\documentclass[aps,prl,reprint,superscriptaddress]{revtex4-2}

\usepackage{graphicx}
\usepackage{calligra}
\usepackage{upgreek}
\usepackage[dvipsnames]{xcolor}
\usepackage{hyperref}
\usepackage{inputenc,caption}
\usepackage{stackrel}
\usepackage{subcaption}
\usepackage{placeins}
\usepackage{amsmath}
\usepackage{amssymb}
\usepackage{soul}

\DeclareMathAlphabet{\mathpzc}{OT1}{pzc}{m}{it}

\newcommand{\be}{\begin{equation}}
\newcommand{\ee}{\end{equation}}
\newcommand{\bea}{\begin{eqnarray}}
\newcommand{\eea}{\end{eqnarray}}
\newcommand{\lb}{\label}

\newcommand{\bu}{{\bf u}}

\newcommand{\bw}{{\bf w}}

\newcommand{\bx}{{\bf x}}

\newcommand{\br}{{\bf r}}

\newcommand{\bR}{{\bf R}}

\newcommand{\grad}{{\mbox{\boldmath $\nabla$}}}
\newcommand{\bdot}{{\mbox{\boldmath $\cdot$}}}

\newcommand{\bcirc}{{\mbox{\boldmath $\circ$}}}

\definecolor{CyanParula}{HTML}{06a4ca}
\definecolor{GreenParula}{HTML}{87bf77}
\definecolor{YellowParula}{HTML}{fec832}

\begin{document}

\setlength{\abovedisplayskip}{8pt}
\setlength{\belowdisplayskip}{8pt}

\title{Non-Gaussian statistics of concentration fluctuations in free liquid diffusion}

\author{Marco Bussoletti}
\affiliation{Department of Mechanical and Aerospace Engineering, Sapienza University of Rome, via Eudossiana 18, 00184 Rome, Italy}
\author{Mirko Gallo}
\affiliation{Department of Mechanical and Aerospace Engineering, Sapienza University of Rome, via Eudossiana 18, 00184 Rome, Italy}
\author{Amir Jafari}
\affiliation{Department of Applied Mathematics and Statistics, The Johns Hopkins University, Baltimore, MD, USA, 21218}
\author{Gregory L. Eyink}
\affiliation{Department of Applied Mathematics and Statistics, The Johns Hopkins University, Baltimore, MD, USA, 21218}
\affiliation{Department of Physics and Astronomy, The Johns Hopkins University, Baltimore, MD, USA, 21218}

\begin{abstract}
We show that the three-point skewness of concentration fluctuations is non-vanishing
in free liquid diffusion, even in the limit of vanishingly small mean concentration gradients. 
We exploit a high-Schmidt reduction of nonlinear Landau-Lifshitz hydrodynamics for a binary fluid, 
both analytically and by a massively parallel Lagrangian Monte Carlo simulation. 
Non-Gaussian statistics result from nonlinear coupling of concentration fluctuations to thermal velocity fluctuations, 
analogous to the turbulent advection of a passive scalar. Concentration fluctuations obey no central limit theorem, 
counter to the predictions of macroscopic fluctuation theory for generic diffusive systems. 
\end{abstract}

\maketitle

\textit{Introduction.} 
The origin of macroscopic hydrodynamics from microscopic molecular dynamics is 
a vexing open problem in physics, with various competing proposals such as
Zwanzig-Mori coarse-graining methods \cite{zubarev1983statistical,espanol2009microscopic}, 
hydrodynamic scaling limit (HSL) \cite{spohn2012large} and macroscopic fluctuation theory 
(MFT) \cite{bertini2015macroscopic}, renormalization group (RG) \cite{forster1977large}
and effective field theory (EFT) \cite{liu2018lectures}, 
or kinetic theory for low-density gases \cite{dorfman2021contemporary}. 
The problem became even more complex with the theoretical discovery \cite{kirkpatrick1982light}
and experimental verification \cite{dezarate2006hydrodynamic} of 
generic spatial long-range non-equilibrium correlations. Such second-order correlations are predicted by linearized fluctuating hydrodynamics \cite{ronis1980statistical}
and have been interpreted in MFT as central limit theorem (CLT) 
corrections to a leading-order law of large numbers. Important topics of on-going 
research 
are the dynamical emergence of these correlations \cite{doyon2023emergence} 
and the higher-order correlations \cite{bertini2015macroscopic,delacretaz2024nonlinear}, 
especially as the latter 
can be measured experimentally in some condensed-matter systems, such as atomic superfluids
\cite{schweigler2017experimental}. Indeed, the universal validity of MFT has been challenged 
by the prediction of non-Gaussian statistics of charge density fluctuations in Dirac metals
based on nonlinear fluctuating hydrodynamics \cite{gopalakrishnan2024non},  with higher statistics 
different from MFT predictions for generic diffusive systems \cite{delacretaz2024nonlinear}.

In this Letter we show that non-Gaussian statistics appear already in a very commonplace 
example: the free diffusion of a solute in a liquid solvent, such as a drop of dye in water. 
Long-range correlations in free liquid diffusion were long ago observed in terrestrial 
experiments by light-scattering \cite{vailati1997giant}. Although quenched at large-scales 
by bouyancy effects on earth, they are observed up to system size in on-going low-gravity space experiments
\cite{croccolo2016shadowgraph,vailati2024perspective}. The concentration
second-order correlations were successfully predicted by linearized fluctuating hydrodynamics 
\cite{vailati1998nonequilibrium} and fluctuations are thus widely expected to be Gaussian \cite{brogioli2016correlations}. 
However, in a landmark work 
\cite{donev2014reversible}, Donev, Fai \& vanden-Eijnden (hereafter, DFV) succeeded to explain 
these fluctuations in a liquid at rest by a nonlinear advection equation 
$\partial_t c+ \bu\bdot\grad c = D_0\triangle c,$
where $c$ is the concentration of the solute, $\bu$ is the thermal velocity field of the solvent 
liquid governed by linearized fluctuating hydrodynamics, and $D_0$ is the bare diffusivity.
In agreement with the EFT interpretation, there is an explicit high-wavenumber cut-off in this model taken 
to be of the order of the radius $\sigma$ of the solute molecule. 
In the limit of  high Schmidt number $\nu/D_0\gg 1$ appropriate to most liquid mixtures, 
DFV showed that this system 
reduces to a version of the {\it Kraichnan model of turbulent scalar advection}
\cite{kraichnan1968small,falkovich2001particles}:
\be \partial_t c+ \bw\bcirc\grad c = D_0\triangle c \lb{ceq2} \ee 
where $\bcirc$ denotes the Stratonovich product and $\bw$ is a Gaussian random velocity
with zero mean and covariance  
$ \langle w_i(\bx,t) w_j(\bx',t') \rangle = R_{ij}(\bx,\bx') \delta(t-t') $
where $\bR=\frac{2k_BT}{\eta}{\bf G}$ with ${\bf G}$ the Greens function of the Stokes operator (Oseen tensor). 
The DFV theory leads 
naturally to a renormalization of the bare diffusivity $D_0$, so that the mean concentration field
satisfies a Fickian diffusion equation 
$\partial_t \bar{c}({\bf x}, t) = D \triangle_{{\bf x}} \bar{c}({\bf x}, t),$ 
in which the diffusivity at macroscopic scales is given by the Stokes-Einstein relation 
$D=k_BT/6\pi \eta\sigma\gg D_0,$ where $T$ is temperature and $\eta=\rho\nu $ the shear viscosity of the solvent. 
DFV 
explained the emergence of long-range correlations 
as closely analogous to a turbulent cascade of an advected scalar. 
Just as in the Kraichnan model \cite{falkovich2001particles}, closed equations hold 
\begin{eqnarray}\nonumber
 &&\partial_t {\mathcal C}_{12}=D (\triangle_{{\bf x}_1} +\triangle_{{\bf x}_2} ){\mathcal C}_{12}
 +{1\over 2} 
{ R}_{ij}(\bx_1,\bx_2)\nabla_{x_1^i}\nabla_{x_2^j} {\mathcal C}_{12} \\\label{C2eq}
&&\; \hspace{50pt} + R_{ij}({\bf x}_1,{\bf x}_2) \nabla_{x_1^i}\bar{c}_1 \nabla_{x_2^j}\bar{c}_2
\lb{C2eq} \end{eqnarray}
for second cumulants ${\mathcal C}_{12}={\mathcal C}(\bx_{1},\bx_{2},t),$ also  
\begin{eqnarray}
&& \partial_t {\mathcal C}_{123}=D\Big[\triangle_{{\bf x}_1}+\triangle_{{\bf x}_2}+\triangle_{{\bf x}_3} \Big]{\mathcal C}_{123}\cr
&& +R_{ij}({\bf x}_1-{\bf x}_2)\nabla_{x_1^i}\nabla_{x^j_2}{\mathcal C}_{123}+R_{ij}({\bf x}_1-{\bf x}_3)\nabla_{x_1^i}\nabla_{x^j_3}{\mathcal C}_{123}\cr
&&\hspace{60pt} +R_{ij}({\bf x}_2-{\bf x}_3)\nabla_{x_2^i}\nabla_{x^j_3}{\mathcal C}_{123}\cr
&&\hspace{10pt}+R_{ij}({\bf x}_1-{\bf x}_2)\Big[ \nabla_{x_1^i}\overline{c_1} \; \nabla_{x^j_2}{\mathcal C}_{23}
+ \nabla_{x^j_2}\overline{c_2} \; \nabla_{x_1^i}{\mathcal C}_{13}
\Big]\cr 
&&\hspace{10pt}+R_{ij}({\bf x}_1-{\bf x}_3)\Big[ \nabla_{x_1^i}\overline{c_1} \; \nabla_{x^j_3}{\mathcal C}_{23} 
+ \nabla_{x^j_3}\overline{c_3} \; \nabla_{x_1^i}{\mathcal C}_{12}
\Big]\cr
&&\hspace{10pt}+R_{ij}({\bf x}_2-{\bf x}_3)\Big[ \nabla_{x_2^i}\overline{c_2} \; \nabla_{x^j_3}{\mathcal C}_{13}
+ \nabla_{x^j_3}\overline{c_3} \; \nabla_{x_2^i}{\mathcal C}_{12}
\Big], \label{C3eq}
\end{eqnarray}
for triple cumulants ${\mathcal C}_{123}={\mathcal C}(\bx_{1},\bx_{2},\bx_{3},t)$ 
of concentration fluctuations, and so forth. In these
equations, gradients of the lower-order cumulants appear as source terms. It was shown recently 
\cite{eyink2024kraichnan} that the DFV model yields second-order correlations 
in the form observed experimentally \cite{vailati1997giant,croccolo2007nondiffusive}:
${\mathcal C}_{12}\propto |\nabla\bar{c}(t)|^2\sigma r_{12}, $
where $\nabla\bar{c}(t)$ is the mean concentration gradient and $r_{12}=|\bx_1-\bx_2|.$

It was pointed out also in \cite{eyink2024kraichnan} that the higher-order cumulants of concentration 
fluctuations should not vanish in the DFV theory, just as they do not in turbulent 
scalar cascades. The structure of the source terms in the closed equations \eqref{C2eq},\eqref{C3eq}, etc. suggests 
that, in general, the $P$th-order cumulant $C_{12...P}\propto (\nabla\bar{c})^P.$ If so, not only are cumulants 
for all $P>2$ non-vanishing, but also when normalized by the $P/2$ power of the second-cumulant 
${\mathcal C}_{12}$ they will remain non-zero in the limit $|\nabla \bar{c}|\to 0.$ Such higher cumulants 
non-dimensionalized by the second cumulant generalize the concepts of skewness and kurtosis and their 
non-vanishing indicates persistent non-Gaussianity as $|\nabla \bar{c}|\to 0.$ A main assumption 
of MFT \cite{bertini2015macroscopic} is that macroscopic diffusion equations 
arise as a ``law of large numbers'' in a hydrodynamic scaling limit \cite{spohn2012large}
where macroscopic gradients become arbitrarily weak on the microscopic scale.
The experimentally observed second cumulants of concentration fluctuations are then interpreted in MFT 
as ``central limit theorem'' corrections, predicted to become Gaussian in the limit
\cite{spohn2012large,bertini2015macroscopic}. 

To address these questions, we consider here the simplest situation \cite{eyink2024kraichnan}
with initial concentration profile 
$c_0({\bf x})=\frac{c_0}{2}\left( 1+{\rm erf}\left(\frac{z}{2\sqrt{D\tau}}\right)\right)$   
in infinite three-dimensional space and gradients nonzero only in the $z$-direction. 
The parameter $\tau$ with units of time is introduced to set the magnitude of the initial concentration 
gradient as $|\nabla c_0|=c_0/\sqrt{4\pi D\tau}.$ Our problem set-up idealizes laboratory experiments \cite{vailati1997giant,croccolo2007nondiffusive} on free diffusion with initial interfaces in 
concentration, but where the spatial domain was finite and gravity appeared. Following DFV, we ignore 
initial local equilibrium fluctuations of concentration, which are orders of magnitude smaller 
than the non-equilibrium fluctuations of interest (see \cite{vailati1997giant}, Fig.~2).
Note that the mean concentration $\bar{c}(\bx,t)$ obtained by evolving
the error function profile under the diffusion equation 
keeps the same form, but 
with $\tau\mapsto t+\tau.$ To probe the non-Gaussian statistics, we will calculate the three-point skewness 
function ${\mathcal S}_{123}={\mathcal C}_{123}/({\mathcal C}_{13}{\mathcal C}_{23}{\mathcal C}_{12})^{1/2}.$ 

We focus first on the early-time, transient regime when non-equilibrium correlations 
begin to emerge. In fact, in the free-decay problem where the mean-gradient decreases 
in time as $|\nabla \bar{c}|\sim c_0/\sqrt{4\pi D(t+\tau)},$ the largest cumulants 
in absolute magnitude appear near the end of this early-time regime \cite{bussoletti2025emergence}
and thus should present the largest experimental signal. This early-time regime was first 
investigated theoretically by DFV \cite{donev2014reversible} whose results for the spectral structure 
function of concentration were confirmed in physical space to have the form \cite{bussoletti2025emergence}:
\bea \lb{larger}
&& C(r_{12}, \theta_{12},Z_{12},t)\simeq {3 c_0^2\over 8\pi} \cdot {\sigma \over r_{12}}(1+\cos^2\theta_{12}) \cr 
&& \quad \times \left[ E_1\left( \frac{Z^2_{12}+\frac{1}{4}z^2_{12}}{2D(t+\tau)}\right) - E_1\left( \frac{Z^2_{12}
+\frac{1}{4}z^2_{12}}{2D\tau}\right) \right],  \cr
&& \qquad \qquad \qquad \qquad Dt\lesssim \sigma^2\ll r^2, 
\lb{C2asympt} 
\eea 
where $r_{12},\theta_{12}$ are spherical coordinates (omitting azimuthal angle) of the separation vector 
$\br_{12}=\bx_1-\bx_2$
of the two points, $Z_{12}=(z_1+z_2)/2$ is the mean of their vertical positions, $z_{12}=r_{12}\cos\theta_{12}$ 
is the vertical coordinate of $\br_{12},$ and  $E_1(x)$ is the exponential integral function: see \cite[\href{https://dlmf.nist.gov/6.6.E1}{(6.6.1)}]{NIST:DLMF}. 
The result \eqref{larger} is derived from \eqref{C2eq} by a short-time, large-$r$ asymptotic expansion  
in which the source term is evaluated with the infinite-space Oseen tensor $G_{ij}=\frac{1}{8\pi r}\left(\delta_{pq}
+\frac{r_ir_j}{r^2}\right)$ and treated as the leading term on the righthand side, the diffusion term 
being regarded as a small perturbation. The formula \eqref{larger} then results from direct time integration 
of the source term. The triple cumulant ${\mathcal C}_{123}$ can be evaluated in the analogous 
asymptotic scheme. With the symmetries of the interface problem, the 
triple cumulant is specified at each time $t$ by the triangle 
in ${\mathbb R}^3$ with vertices $\bx_1,\bx_2,\bx_3$ or, more geometrically, by the lengths of its three sides
$r_{13},$ $r_{23},$ $r_{12},$ by two Euler angles $\beta,$ $\gamma$ specifying its orientation, and by the
vertical position $Z=z_3.$ See SM, \S A. The asymptotic solution for short times $Dt\lesssim \sigma^2$ 
and large triangles $\min_{p,q} R_{pq} \gg \sigma$ is obtained by integrating in time
the six source terms on the rightthand side of \eqref{C3eq}, treating the diffusion terms as a perturbation.
Although the calculation is straightforward in principle, the resulting expression for ${\mathcal C}_{123}$ 
is quite lengthy. See SM, \S A for details of the derivation and the full analytical result.

Especially relevant to the validity of a central limit theorem 
is ${\mathcal C}_{123}$ in the weak-gradient limit $|\nabla c_0|\to 0$ or, equivalently, 
$\tau\to\infty.$ The analytical expressions simplify in that limit so that, for example,
the result analogous to \eqref{larger} for general particle pair $p,q$ becomes \cite{bussoletti2025emergence}
\bea 
&& {\mathcal C}(r_{pq}, \theta_{pq},Z_{pq},t)\simeq {3\over 2} \sigma Dt 
{|\nabla c_0|^2\over r_{pq}} (1+\cos^2\theta_{pq}), \cr 
&& \qquad \qquad \qquad  D\tau\gg \max\{r^2_{pq},Z^2_{pq}\}\gg \sigma^2\gtrsim Dt
\lb{C2asympt-large} \eea 
The latter conditions for all pairs $p,q$ guarantee not only that $\sigma |\nabla c_0|/c_0\ll 1$ but also 
that the triangle with vertices $\bx_1, \bx_2, \bx_3$ lies entirely in the interface
region 
where the mean concentration gradient $\sim |\nabla c_0|$
is nearly a space-time constant. The time-integrals of the six forcing terms on the righthand side of 
\eqref{C3eq} 
can be computed in the same manner, yielding 
${\mathcal C}_{123}\propto |\nabla c_0|^3 (\sigma Dt)^2/r^3$ 
(where $r$ is some combination of $r_{13},$ $r_{23},$ $r_{12}$ 
depending on $\beta,$ $\gamma$), 
so that skewness ${\mathcal S}_{123}\propto (\sigma D t/r^3)^{1/2}$ is growing in time and 
independent of $|\nabla c_0|,$ even as $|\nabla c_0|\to 0.$ 
See the End Matter for a sketch of the calculation and SM, \S A for full details.  

\begin{figure*}[t!]
\includegraphics[width=1.5\columnwidth]{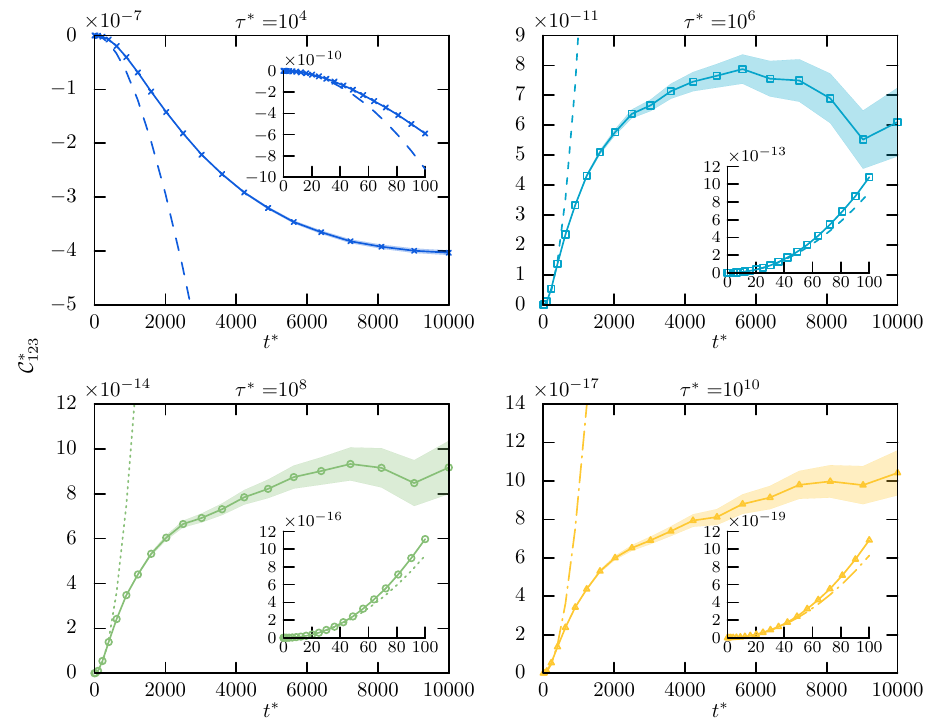}
\caption{\footnotesize {Triple cumulants for the vertical equilateral triangle with side lengths $r^*=50$ and for initial mean concentration profiles with $\tau^*=10^4$, $10^6$, $10^8$ and $10^{10}$ going from left to right, top to bottom, respectively. Solid lines with symbols and shades show the numerical results and corresponding errors from the Lagrangian Monte Carlo computation. Dashed lines correspond to the analytical expression in SM, \S A, valid asymptotically for short times, as highlighted by the insets. For 
corresponding results on the combined second cumulant $({\mathcal C}_{13}{\mathcal C}_{23}{\mathcal C}_{12})^{1/2}$ used to calculate the three-point skewness, see End Matter, Fig.\ref{C_2 r=50}.}} 
\label{Fig1}
\end{figure*}

These conclusions may be confirmed and extended by an independent evaluation
of the concentration cumulants with a Lagrangian Monte Carlo method, first employed for 
passive scalars in the Kraichnan model of turbulent advection 
\cite{frisch1998intermittency,frisch1999lagrangian,gat1998anomalous}
and recently applied by us to liquid mixing \cite{bussoletti2025emergence}. In this method,
the DFV model \eqref{ceq2} is solved for $D_0=0$ at the three points $\bx_p,$ $p=1,2,3$
in one realization of the thermal velocity $\bw$ by setting $c({\bf x}_p,t)=c_0({\boldsymbol \xi}_p(t))$
where $d{\boldsymbol \xi}_p/dt={\bf w}_p,$ ${\boldsymbol \xi}_p(0)=\bx_p,$ $p=1,2,3$ with 
${\boldsymbol {\mathcal W}}=({\bf w}_1,{\bf w}_2,{\bf w}_3)$ a 9-dimensional Gaussian white-noise with covariance 
\begin{equation}
   \langle w_{pi}(t)w_{qj}(t')\rangle_{\boldsymbol\xi}=R_{ij}({\boldsymbol \xi}_p(t),{\boldsymbol \xi}_q(t))\delta(t-t')
   \,,
   \label{eq:R}
\end{equation}
for $p,q = 1,...,3$ and $i,j=1,...,3.$ By repeating this numerical integration for independent 
realizations ${\boldsymbol {\mathcal W}}^{(n)},$ $n=1,...,N,$ the triple cumulant 
can then be evaluated by an empirical average over samples
\be {\mathcal C}_{123} \simeq \frac{1}{N} \sum_{n=1}^N c^{\prime(n)}({\bf x}_1,t)c^{\prime(n)}({\bf x}_2,t) 
c^{\prime(n)}({\bf x}_3,t) 
\lb{eq:sample} \ee 
where $c^{\prime(n)}({\bf x},t)=c^{(n)}(\bx,t)-\bar{c}(\bx,t).$
This scheme yields the exact cumulants without restriction to short times and large-$r$ but 
it is subject to slow Monte Carlo convergence and can require a large number $N$ of samples.
In the original application to turbulent advection 
\cite{frisch1998intermittency,frisch1999lagrangian,gat1998anomalous} a number of samples 
$N\simeq 10^7$ sufficed, but our problem is much more demanding. First, the correlations 
between thermal velocity fluctuations decay with distance, opposite to the growing correlations 
for turbulent velocities. Second, we consider a problem of free decay rather than a 
statistical steady state as in the turbulence studies 
\cite{frisch1998intermittency,frisch1999lagrangian,gat1998anomalous}, and 
the decaying mean concentration field generates higher cumulants with absolute magnitude
decreasing in time. 
For these reasons, accurate results for 
our case required $N\simeq 10^{14}$ samples, which we only achieved by a CUDA-MPI 
implementation designed to leverage the massive parallel-processing capabilities of GPU's. 
For more details on the numerics, see \cite{bussoletti2025emergence} and SM, \S B. All results 
are calculated in dimensionless form, with $r^*=r/\sigma,$ $t^*=2Dt/\sigma^2,$ $c^*=c/c_0$

\begin{figure}[t]
\hspace{-30pt} 
\includegraphics[width=1.1\columnwidth]{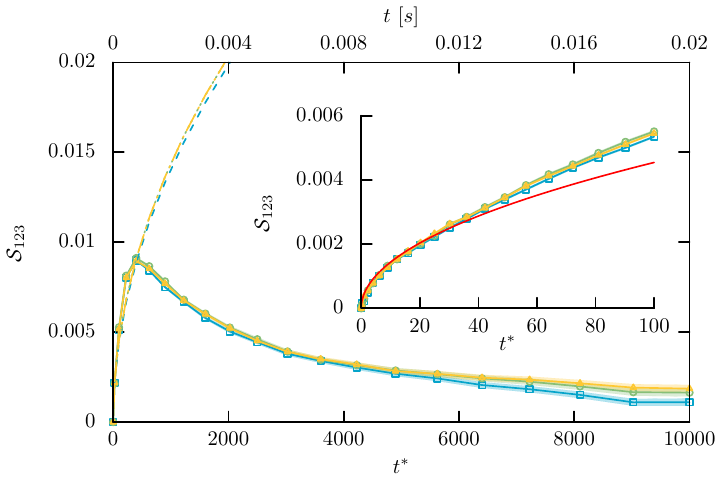}
\caption{\footnotesize {Three-point skewness for the vertical equilateral triangle with side lengths $r^*=50$ and small concentration gradients (colors and symbols are coded as in Fig.~\ref{Fig1} for $\tau^*=10^6$ - cyan squares (\textcolor{CyanParula}{${\boldsymbol \square}$}), $10^8$ - green circles  (\textcolor{GreenParula}{${\boldsymbol \circ}$}), $10^{10}$ - yellow triangles (\textcolor{YellowParula}{${\boldsymbol \triangle}$}). 
The result for $\tau^*=10^4$ can be found in SM, \S C.) Solid lines with symbols and shades represent the numerical Lagrangian results and Monte Carlo errors, while dashed lines show the analytical prediction. The times $t$ in seconds on the upper axis are calculated with reference
values $\sigma=10$ nm and and $D=2.2\times 10^{-5}\,{\rm mm}^2/{\rm sec}$ for which the maximum skewness 
is achieved in about $4$ ms. Short-time behavior is zoomed in the inset, with the red line 
 (\textcolor{red}{${\boldsymbol -}\!\!{\boldsymbol -}$})
showing the short-time asymptotic prediction for vanishing concentration gradient.}}
\label{Fig2}
\end{figure}

We present in Fig.~\ref{Fig1} the triple cumulant ${\mathcal C}_{123}^*$ 
at three points forming an equilateral triangle with sidelength 
$r_{13}^*=r_{23}^*=r_{12}^*=50$ and lying in a plane perpendicular to $z=0$
$(\gamma=0,\beta=\pi/2).$ (Results for another value $r^*=100$ are presented 
in SM, \S C, which lead to the same conclusions.) 
The Lagrangian numerical results for four values 
of $\tau^*=10^4-10^{10}$ are seen generally to grow in magnitude with $t^*$ 
and then roughly to saturate, within increasing Monte Carlo error. They  
agree well with the analytic predictions for shorter $t^*,$
as seen especially in the insets. The non-vanishing triple cumulants demonstrate 
non-Gaussianity of the concentration statistics. Note the intriguing change in 
sign between $\tau^*=10^4$ and $\tau^*=10^6.$ For $\tau^*>10^6,$ however, the 
triple cumulants appear to become qualitatively similar and 
simply decreasing in magnitude with increasing $\tau^*,$ consistent 
with the estimate ${\mathcal C}_{123}\propto |\nabla c_0|^3.$ The latter 
scaling is confirmed by the plot of the 3-point skewness ${\mathcal S}_{123}$ 
in Fig.~\ref{Fig2}, in which the results for the three largest $\tau$-values,
differing over four orders of magnitude, essentially collapse. 
Note that the skewness from the Lagrangian Monte Carlo also agrees well with the 
asymptotic prediction at short-$t^*$ (inset) but thereafter attains a maximum 
value of about 0.01 and then slowly decreases in $t^*$, possibly to a 
non-zero quasi-steady value. Crucially, the skewness is essentially 
$\tau$-independent, not only in the early-time regime but over the entire 
$t^*$-range, and there is no central limit theorem with zero skewness
in the limit of vanishingly small gradients,
$\nabla c_0\to 0$ or $\tau\to\infty.$ Figure \ref{Fig2} constitutes 
the main result of this Letter. 

We have thus shown that nonlinear Landau-Lifschitz fluctuating hydrodynamics 
for a binary mixture, in the high-Schmidt limit appropriate to liquid diffusion, 
implies non-Gaussian statistics of concentration fluctuations. An experimental 
test of these predictions might be feasible by light-scattering methods 
\cite{schaefer1972light,lemieux1999investigating}.
We have considered 
free diffusion of an initial inhomogeneous concentration field, but it would be interesting 
to investigate 
also non-equilibrium steady states.
Beyond correlations,
other statistics could show turbulent-like, non-Gaussian features such as fat tails 
in probability density functions \cite{shraiman1994lagrangian}. Note, however, that 
this can occur only for multi-point variables such as concentration increments, 
as the PDF of the single-point concentration indeed becomes Gaussian as $\nabla c_0\to 0:$
see End Matter. Bouyancy effects 
of gravity that are known to quench long-range concentration correlations can be incorporated
into the DFV model \cite{donev2014reversible} and it is worth understanding 
their influence on non-Gaussianity. In low-gravity environments, by contrast, thermal noise 
effects are observed to propagate to the system size and thus non-Gaussian statistics 
occurring in liquid diffusion may have critical importance for space exploration 
\cite{vailati2024perspective}.

Our results contradict the na\"ive expectations of MFT that statistics of fluctuations 
should become Gaussian in the limit of weak gradients. 
These predictions of MFT are rigorous central limit theorem results for simple toy models of 
particle diffusion,
such as Kawasaki lattice gases, in a suitable `hydrodynamic scaling limit'' (HSL)
\cite{spohn2012large,bertini2015macroscopic}. 
However, it is not at all clear how to formulate a ``hydrodynamic scaling limit'' for the problem of 
liquid diffusion and recent work has shown that the HSL, even when mathematically applicable 
in principle, may be physically unattainable \cite{bandak2022dissipation,bell2022thermal}. 
Furthermore, the DFV model accurately predicts other effects missed by MFT, such as
renormalization of the bare diffusivity \cite{donev2014reversible,brogioli2000giant} 
that is commonly observed in
molecular dynamics studies \cite{donev2011enhancement,celebi2021finite}. The latter renormalization
as well as the non-Gaussian statistics calculated in the present study are due to nonlinear 
coupling of the diffusive mode with the momentum mode, which is a general phenomenon 
predicted by renormalization group \cite{forster1977large} but not included in MFT analyses 
of diffusion \cite{bertini2015macroscopic}. It was a fundamental insight of the Kraichnan model 
that relative advection even by a Gaussian velocity field, such as thermal velocity 
fluctuations in this study, will produce non-Gaussian multi-point statistics for the advected scalar \cite{kraichnan1968small,falkovich2001particles}. 
The non-Gaussian diffusive statistics predicted in Dirac fluids arise from a similar mechanism \cite{gopalakrishnan2024non}. 
The magnitude of the skewness that we observe is not large (only $~0.01$ at maximum in the 
case considered in Fig.\ref{Fig2}) but non-vanishing as $\nabla c_0\to 0,$
indicating persistent nonlinear coupling of concentration and velocity fluctuations. 
We have thus confirmed and extended the claims of earlier work that ``non-equilibrium 
fluctuations do not represent merely a perturbation of a macroscopic state'' 
\cite{brogioli2000diffusive,donev2014reversible,cerbino2015dynamic}. 
Our results imply that MFT and HSL cannot be a universally valid explanation for origin 
of hydrodynamic behavior, not even for the familiar example of liquid diffusion. 

\textit{Acknowledgements.} 
We are grateful to B. Doyon and H. Spohn for discussions of this work.
This research is supported by an ERC grant (ERC-STG E-Nucl. Grant agreement ID: 101163330) (PI M. Gallo).
Support is acknowledged also from the Sapienza Funding Scheme ``Avvio alla Ricerca - Tipo 2", project No. AR22419078AD8186 (PI M. Bussoletti). Computational resources were made available from ICSC-Italian Research Center on High Performance Computing, Big Data, and Quantum Computing under ``MDR-TP - Spoke 6". We also acknowledge support for computational resources from CINECA under the ISCRA iniative, relative to the ISCRA-B D-RESIN (PI M. Gallo) and ISCRA-C EMADON (PI M. Bussoletti) projects. 
G. Eyink thanks the Department of Physics of the University of Rome ‘Tor Vergata’ for hospitality while this work was begun and acknowledges support from the European Research Council (ERC) under the European Union’s Horizon 2020 research and innovation program (Grant Agreement No. 882340). {\it Funded by the European Union}. Views and opinions expressed are however those of the authors only and do not necessarily reflect those of the European Union or the European Research Council Executive Agency. Neither the European Union nor the granting authority can be held responsible for them. 

\textit{Data Availability.} 
The data that support the findings of this study will be available 
through a public repository on the Zenodo platform.

\bibliography{PRLfluctuations.bib}

\clearpage

\appendix*
\begin{center}
\section*{End Matter}
\end{center} 

\subsection*{Sketch of analytical computation of triple cumulants.}

\begin{figure}[h!]
\includegraphics[width=7cm]{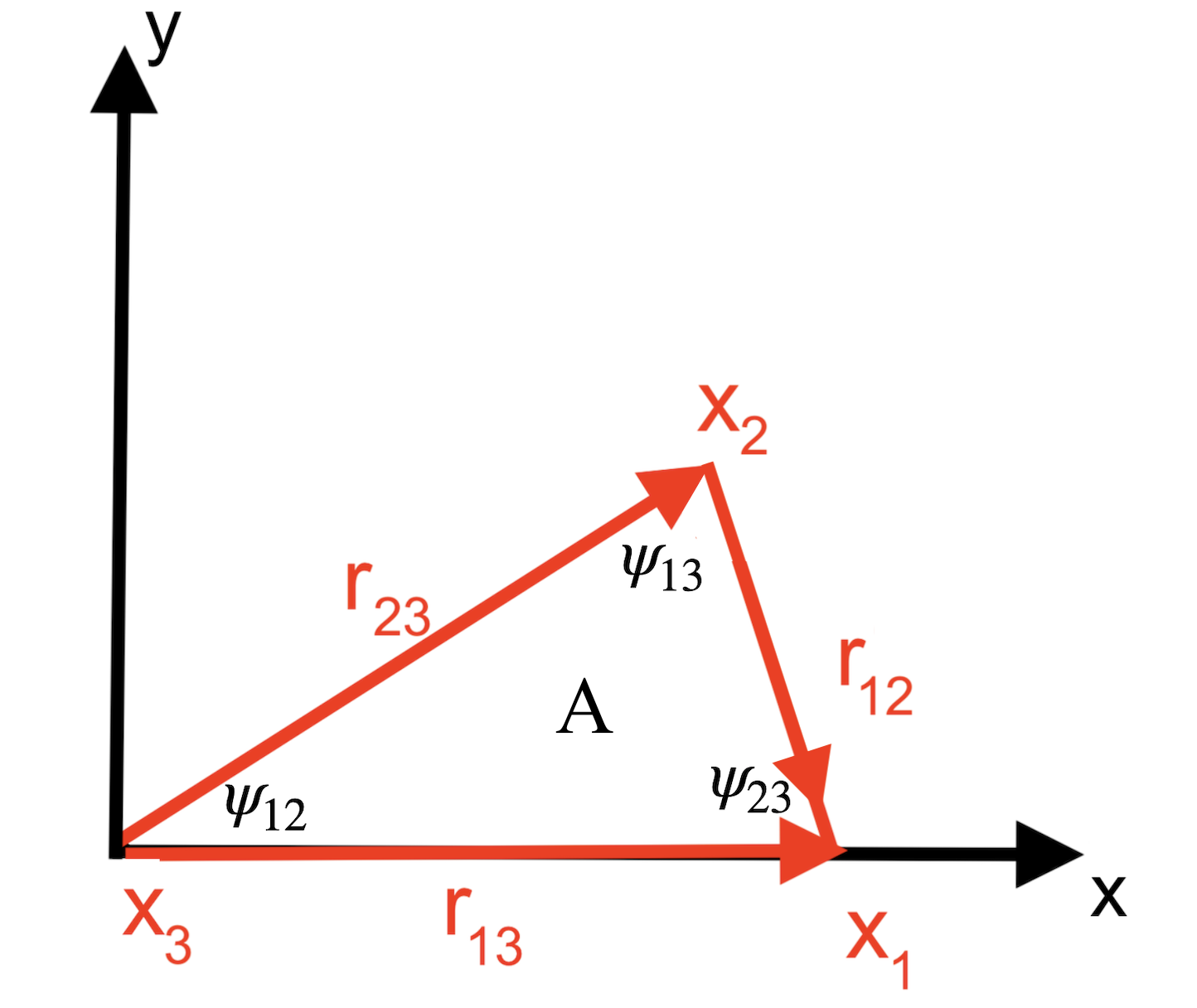}
\caption{\footnotesize {Illustration of the triangle formed by vertices ${\bf x}_1,$ ${\bf x}_2,$ ${\bf x}_3,$ in reference position in the horizontal plane.}} 
\label{triangle}
\end{figure}

We give here a brief sketch of the calculation of the triple cumulant 
of the concentration fluctuations, asymptotically for large separations and for short times. 
We refer to Figure \ref{triangle} for a sketch of the three points in the triple 
cumulant, forming the vertices of a triangle in the plane perpendicular to the mean gradient. 
The triangle is shown in the reference position that we adopt, before rotation by Euler angles 
$\beta,$ $\gamma$ into general orientation. The side lengths $r_{13},$  $r_{23},$  $r_{12}$
and the interior angles $\psi_{13},$ $\psi_{23},$ $\psi_{12}$ specify the geometry of the  
triangle. The source terms on the righthand side of the equation \eqref{C3eq} for the triple 
cumulant that arise from gradients of the mean and second cumulant of the concentration 
depend on both the geometry of the triangle and also on its orientation and displacement 
along the gradient. The latter dependencies appear in the second cumulant ${\mathcal C}_{pq}$
through the azimuthal angle $\theta_{pq}$ and the mean displacement $Z_{pq},$
which ultimately can all be calculated knowing the geometry, orientation and displacement 
of the triangle. Using these variables and the symmetries of the problem, the six source
terms can all be calculated; for example, the first one, $F_{123} = R_{ij}({\bf x}_1-{\bf x}_2)\nabla_{x_1^i}\overline{c_1} \; \nabla_{x^j_2}{\mathcal C}_{23},$ becomes
\begin{eqnarray*}\lb{term1} 
 && F_{123} = \frac{k_BT}{4\pi\eta r_{12}} \bar{c}'(z_1) \Bigg\{ (\cos\theta_{23}	-\cos\psi_{13} \cos\theta_{12}) \frac{\partial{\mathcal C}_{23}}{\partial r_{23}} \cr 
&&  + \frac{\sin^2\theta_{23}+\cos\theta_{12}(\cos\theta_{12}+\cos\psi_{13}\cos\theta_{23})}{r_{23}} \frac{\partial{\mathcal C}_{23}}{\partial (\cos\theta_{23})} \cr
&&  \hspace{50pt} + \frac{1}{2} (1+\cos^2\theta_{12}) \frac{\partial{\mathcal C}_{23}}{\partial Z_{23}} \Bigg\}
\end{eqnarray*} 
and similarly for all six source terms. 

Following the asymptotic scheme outlined in the main text for the second cumulant, the leading 
order contribution to the triple cumulant for large separations and short times is obtained 
simply by integrating in time the contributions from these six source terms. We here 
discuss only the simplifications that occur in the time $\tau\to\infty$. First, 
the mean concentration gradient becomes $t$-independent to leading order:
$\nabla \bar{c} \sim \nabla c_0\sim c_0/\sqrt{4\pi D\tau}.$ In that case, one 
must evaluate time-integrals only of the derivatives of the second cumulants. These are 
easily evaluated for large-$\tau$ from \eqref{C2asympt-large} to be 
\begin{eqnarray*}
&& \int_0^t \frac{\partial {\mathcal C}}{\partial r_{pq}}(r_{pq},\cos\theta_{pq},Z_{pq},s)\, ds \cr 
&& \hspace{50pt} \sim -\frac{3}{4}|\nabla c_0|^2 \frac{\sigma}{r^2_{pq}}(1+\cos^2\theta_{pq}) D t^2
\end{eqnarray*}
and 
\begin{eqnarray*} 
&& \int_0^t \frac{\partial {\mathcal C}}{\partial (\cos\theta_{pq})}(r_{pq},\cos\theta_{pq},Z_{pq},s)\, ds \cr 
&& \hspace{50pt} \sim \frac{3}{2}|\nabla c_0|^2 \frac{\sigma}{r_{pq}} \cos\theta_{pq} D t^2
 \end{eqnarray*} 
 with other contributions asymptotically smaller. Combining the contributions from 
 the six source terms and recalling that $D=k_BT/6\pi\eta\sigma$, one obtains 
${\mathcal C}_{123}\propto |\nabla c_0|^3 (\sigma Dt)^2/r^3$ as claimed in the main text. 
It follows from this result that the 3-point skewness ${\mathcal S}_{123}$ 
will remain finite in the limit of weak gradients or $\tau\to\infty,$ at least 
for the large-separation, short-time regime. See SM, \S A for complete details 
of the calculation and also a short Matlab code to evaluate the triple cumulant as 
a function of time in this regime, for any choice of the three points. 

\subsection*{Numerical results for second cumulants.}

\begin{figure*}[htpb]
\includegraphics[trim={0 0 0 0},clip,width=1.5\columnwidth]{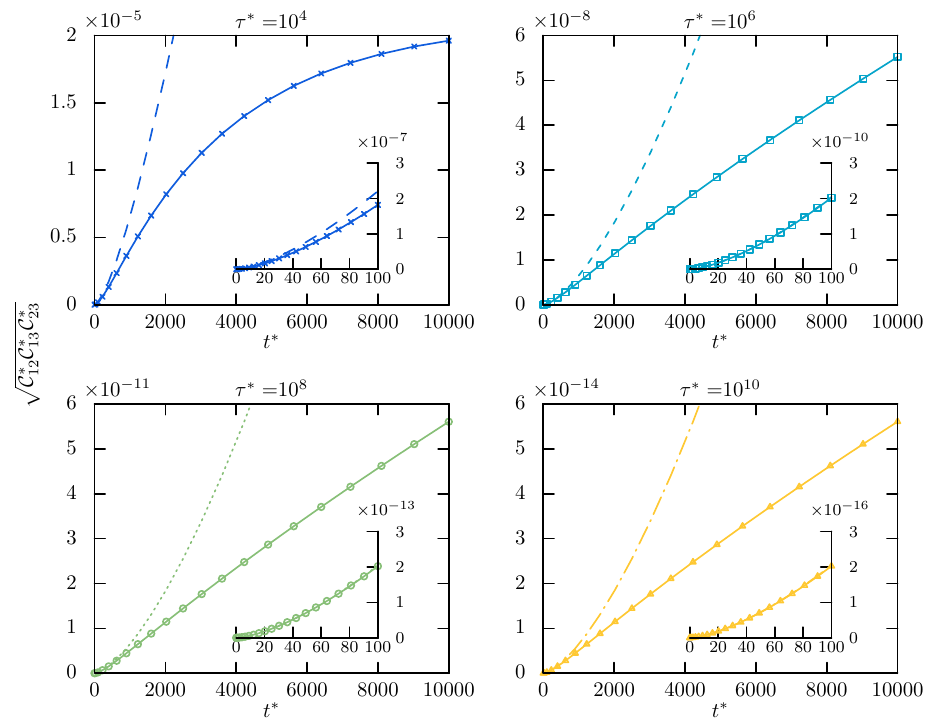}
\caption{\footnotesize {Combined second cumulants $({\mathcal C}_{13}{\mathcal C}_{23}{\mathcal C}_{12})^{1/2}$ 
for the vertical equilateral triangle with side lengths $r^*=50$ and for initial mean concentration profiles with $\tau^*=10^4$, $10^6$, $10^8$ and $10^{10}$ going from left to right, top to bottom, respectively. Solid lines with symbols and shades represent numerical results  and dashed lines the analytical prediction. Monte Carlo error is too small to be visible at this scale.}} 
\label{C_2 r=50}
\end{figure*}

To obtain the Lagrangian Monte Carlo results for the 3-point skewness ${\mathcal S}_{123}$ 
plotted in Fig.~\ref{Fig2}, not only are the results required for the triple cumulant ${\mathcal C}_{123}$ 
plotted in Fig.~\ref{Fig1} but also required are the second cumulants ${\mathcal C}_{13},$ ${\mathcal C}_{23},$ 
${\mathcal C}_{12}.$ The results for the second cumulant from the same Lagrangian Monte Carlo 
scheme were presented in our earlier paper \cite{bussoletti2025emergence} and these were compared
there with the large-separation, short-time approximation \eqref{C2asympt}. However, for completeness,
we present here a similar comparison for the particular combination 
$({\mathcal C}_{13}{\mathcal C}_{23}{\mathcal C}_{12})^{1/2}$ that appears in our definition 
of the three-point skewness. In Figure \ref{C_2 r=50} we show this combination for the same 
three points and for the same $\tau$-values as that for the triple cumulants plotted in Fig.~\ref{Fig1}. 
An interesting observation is that the Monte Carlo errors are much smaller 
for the second cumulants than for the triple cumulants. In fact, it is these small error bars 
which, by standard propagation of error formulas, led to the relative smaller errors in the 
3-point skewness in Fig.~\ref{Fig2} relative to the triple cumulant in Fig.~\ref{Fig1}. 
We have also plotted in Fig. \ref{C_2 r=50} the predictions of the large-separation, 
short-time asymptotics. It can be seen especially in the insets of the figures, which zoom into the 
short-time region, that the agreement between the analytical and numerical results is excellent. 
This close agreement provides a good check on both methods. 

\onecolumngrid 
\twocolumngrid

\vspace{10pt} 
\subsection{1-point PDF's of concentration fluctuations.}

Although it is not immediately obvious how to calculate the probability density function (PDF) 
of concentration fluctuations at multiple space points, it is easy in our model 
for $D_0=0$ to evaluate the PDF at a single point by using the fact that $c({\bf x},t)=
c_0({\boldsymbol \xi}(t))$ where ${\boldsymbol \xi}(0)={\bf x}$ and $
d{\boldsymbol \xi}/dt={\bf w}$ is a 
white-noise process with covariance
$\langle w_i(t) w_j(t')\rangle = 2D\delta_{ij} \delta(t-t').$
Since our non-dimensionalization corresponds to taking $2D=1,$ this gives 
$c^*(z^*,t^*)=c_0^*(z^*+\sqrt{t^*}N)$ in dimensionless variables, 
with $N$ a standard normal random variable. From our adopted initial concentration profile 
$c_0^*(\zeta^*)=\frac{1}{2}\left( 1+{\rm erf}\left(\frac{\zeta^*}{\sqrt{2\tau^*}}\right)\right),$
it is then straightforward to find the PDF of the concentration $c^*(z^*,t^*).$

Because $c_0^*(\zeta^*)$ is a nonlinear function, this random variable is in general 
non-normal. For example, when $t^*=\tau^*,$ the theory of inverse transform sampling 
implies that $c^*(0,t^*)$ has a uniform distribution on the unit interval $[0,1]$
of concentration values and for increasing 

\noindent 
$$\,$$
$t^*>\tau^*$ its PDF becomes more peaked 
at values $c^*=0$ and $c^*=1$. On the other, in the opposite limit of weak concentration 
gradients, $\tau^*\gg t^*, (z^*)^2/2,$ the result 
$$ c_0^*(\zeta^*) \simeq \frac{1}{2}\left(1+ \left(\frac{2}{\pi\tau^*}\right)^{1/2}\zeta^*\right)$$
implies that 
$$ c^*(z^*,t^*) \simeq \frac{1}{2}\left(1+ \left(\frac{2t^*}{\pi\tau^*}\right)^{1/2}N\right)$$
and converges to a Gaussian random variable with mean $\frac{1}{2}$ and standard deviation 
$\propto \nabla c_0^*$
It should be noted that a similar result holds also for $D_0>0,$ since this amounts 
simply to replacing $D\mapsto D+D_0.$

Thus, the single-point statistics of the concentration indeed become Gaussian in the limit 
$\nabla c_0\to 0.$ This fact underlines the non-triviality of our results for multi-point 
statistics, which remain non-Gaussian in this same limit. 

\end{document}


\title{Supplemental Material on ``Non-Gaussian statistics of concentration fluctuations in free liquid diffusion''}
\author{Marco Bussoletti}
\affiliation{Department of Mechanical and Aerospace Engineering, Sapienza University of Rome, via Eudossiana 18, 00184 Rome, Italy}
\author{Mirko Gallo}
\affiliation{Department of Mechanical and Aerospace Engineering, Sapienza University of Rome, via Eudossiana 18, 00184 Rome, Italy}
\author{Amir Jafari}
\affiliation{Department of Applied Mathematics and Statistics, The Johns Hopkins University, Baltimore, MD, USA, 21218}
\author{Gregory L. Eyink}
\affiliation{Department of Applied Mathematics and Statistics, The Johns Hopkins University, Baltimore, MD, USA, 21218}
\affiliation{Department of Physics and Astronomy, The Johns Hopkins University, Baltimore, MD, USA, 21218}

\maketitle

\section{Short-time, large-$r$ asymptotic solution for triple cumulant}

\begin{figure}
\includegraphics[width=7cm]{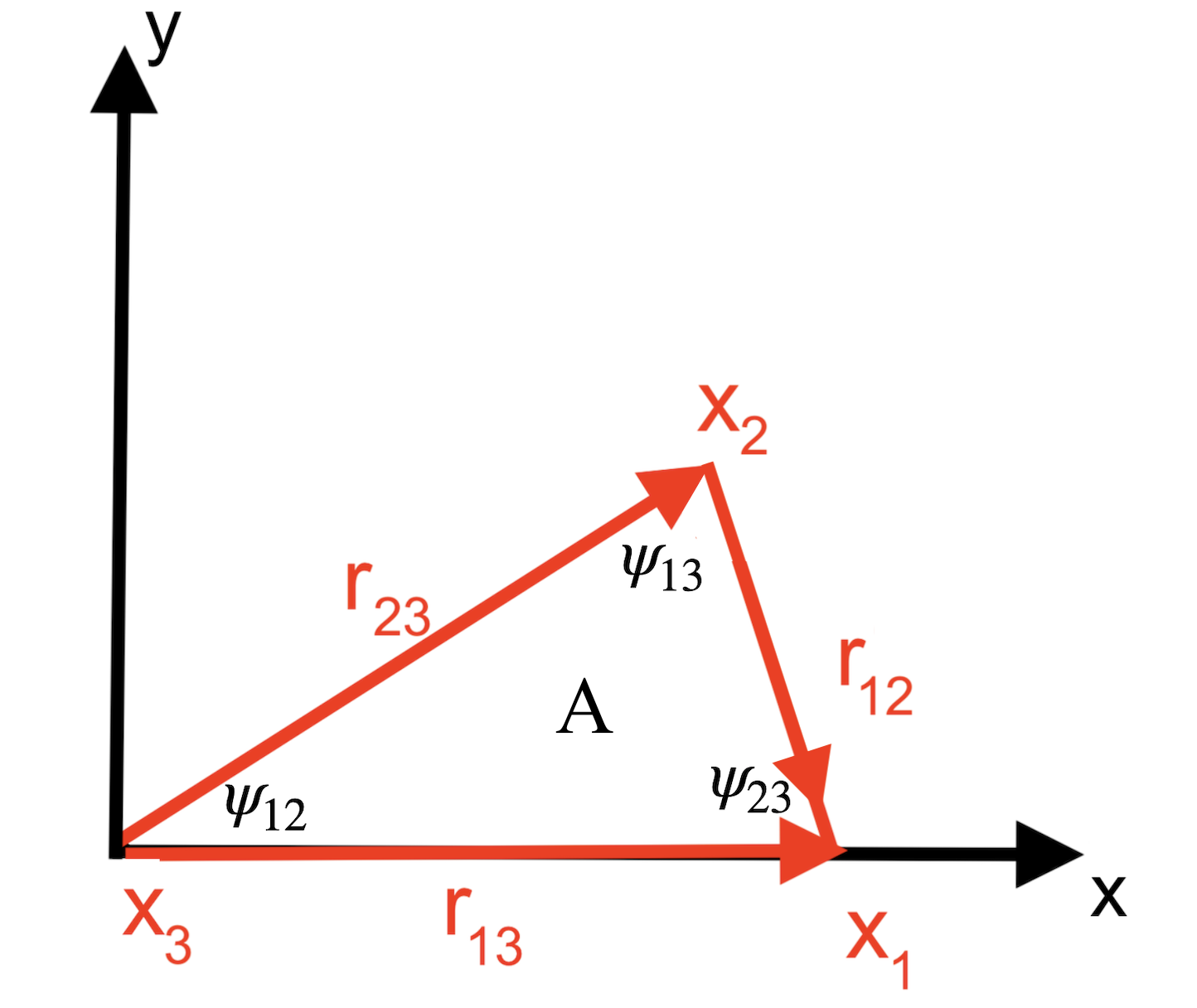}
\caption{\footnotesize {Illustration of the triangle formed by vertices ${\bf x}_1,$ ${\bf x}_2,$ ${\bf x}_3,$ in reference position in the horizontal plane.}} 
\label{triangle}
\end{figure}

We first describe the variables used to represent ${\mathcal C}_{123}$, in terms of the geometry of the triangle 
with vertices $\bx_1,\bx_2,\bx_3.$
We take as reference the triangle with sides $r_{12}, r_{13}, r_{23}$ and interior angles $\psi_{12}, \psi_{13}, \psi_{23}$ 
located in the first quadrant of the $xy$-plane with vertex 3 at the origin and side $\br_{13}$ along the $x$-axis. 
See Figure~\ref{triangle}. Applying rotation $R(\alpha, \beta, \gamma)=R_z(\alpha)
R_x(\beta)R_z(\gamma)$ around point 3 and then translating point 3 by $(X,Y,Z),$ the general position and orientation of the triangle as a rigid body in space is specified by $9$ variables: $3$ side-lengths $r_{12}, r_{13}, r_{23}$, $3$ Euler angles $\alpha, \beta, \gamma,$ and 
3 coordinates $(X,Y,Z)$ of vertex 3. The number of independent variables is reduced from $9$ to $6$ by using translational symmetry 
along the planar interface (eliminating $X,Y$) and rotational symmetry around the gradient direction (eliminating 
$\alpha$). Transforming Cartesian variables ${\bf x}_1, {\bf x}_2, {\bf x}_3$ to these new variables, therefore, the $3$rd order cumulant ${\mathcal C}_{123}$ can be written entirely in terms of  $r_{12}, r_{23}, r_{13}, \beta, \gamma, Z.$ The shape variables are subject to the triangle inequalities $r_{pq}\leq r_{ik}+r_{jk}$ for $i,j,k$
any cyclic permutation of $1,2,3.$ It is thus more convenient to use unconstrained variables $r_{13}, r_{23}, \psi_{12}$
to specify triangle shape and recover the others from the trigonometric laws of cosines and sines:  
\bea 
r_{12}=(r_{13}^2+r_{23}^2-2r_{13}r_{23}\cos(\psi_{12}))^{1/2}; \quad 
\psi_{23}=\arcsin(r_{23}\sin(\psi_{12})/r_{12}); \quad
\psi_{13}=\arcsin(r_{13}\sin(\psi_{12})/r_{12}) \eea 
Triangle orientation is specified by rotating 
coordinate vectors $\hat{{\bf x}},$ $\hat{{\bf y}},$ $\hat{{\bf z}},$ by $R_x(\beta)R_z(\gamma)$ into a new orthogonal system
\bea 
{\bf l}=(\cos\gamma, \cos\beta \sin\gamma, \sin\beta\sin\gamma)^\top; \quad
{\bf m}=(-\sin\gamma, \cos\beta\cos\gamma, \sin\beta\cos\gamma)^\top; \quad 
{\bf n}=(0, -\sin\beta, \cos\beta)^\top,  \eea 
which then gives the triangle side vectors ${\bf r}_{13},$ ${\bf r}_{23},$ ${\bf r}_{12},$ in general position:
\bea 
\br_{13}=r_{13}{\bf l}; \quad 
\br_{23}=r_{23}(\cos\psi_{12}{\bf l}+\sin\psi_{12}{\bf m}); \quad 
\br_{12}=r_{12}(\cos\psi_{23}{\bf l}-\sin\psi_{23}{\bf m}). \lb{sides} \eea 

The variables used to represent the second cumulant ${\mathcal C}_{pq}$ for the two points ${\bf x}_p,$ ${\bf x}_q$ 
are the radial distance $r_{pq},$ the polar angle $\theta_{pq},$ and the mean height $Z_{pq}=\frac{1}{2}(z_p+z_q).$
The forcing terms in equation (4) of the Main Text for ${\mathcal C}_{123}$ involve the gradients of the second 
cumulants and thus we must relate its variables to those that we have adopted for ${\mathcal C}_{123}$.  
The $z-$coordinates of the triangle side vectors can be represented also by the polar angles relative to vector $\hat{{\bf z}}$
\be z_{13}=r_{13}\cos\theta_{13};\ z_{23}=r_{23}\cos\theta_{23};\ z_{12}=r_{12}\cos\theta_{12}, \ee 
so that comparing with equation \eqref{sides} yields: 
\bea 
\cos \theta_{13}= \sin\beta\sin\gamma; \quad
\cos\theta_{23}=\sin\beta \sin(\gamma+\psi_{12}); \quad 
\cos\theta_{12}=\sin\beta\sin(\gamma-\psi_{23}); \eea  
Recalling $Z=z_3,$ the mean vertical heights of the triangle legs are then given by 
\be  Z_{13}=Z+\frac{1}{2}z_{13};\quad Z_{23}=Z+\frac{1}{2}z_{23};\quad Z_{12}=Z+\frac{1}{2}(z_{13}+z_{23}) \ee 

The six forcing or source terms that appear on the right side of equation (4) of the Main Text for ${\mathcal C}_{123}$ 
can be obtained by straightforward but lengthy computations. We give only the final results: 
\begin{eqnarray}\lb{term1} 
 && F_{123} = \frac{k_BT}{4\pi\eta r_{12}} \bar{c}'(z_1) \Bigg\{ (\cos\theta_{23}	-\cos\psi_{13} \cos\theta_{12}) \frac{\partial{\mathcal C}_{23}}{\partial r_{23}} 
 + \frac{\sin^2\theta_{23}+\cos\theta_{12}(\cos\theta_{12}+\cos\psi_{13}\cos\theta_{23})}{r_{23}} \frac{\partial{\mathcal C}_{23}}{\partial (\cos\theta_{23})} \cr
&&  \hspace{320pt} + \frac{1}{2} (1+\cos^2\theta_{12}) \frac{\partial{\mathcal C}_{23}}{\partial Z_{23}} \Bigg\}
\end{eqnarray} 
\begin{eqnarray}\lb{term1} 
 && F_{213} = \frac{k_BT}{4\pi\eta r_{12}} \bar{c}'(z_2) \Bigg\{ (\cos\theta_{13}	+\cos\psi_{23} \cos\theta_{12}) \frac{\partial{\mathcal C}_{13}}{\partial r_{13}} 
 + \frac{\sin^2\theta_{13}+\cos\theta_{12}(\cos\theta_{12}-\cos\psi_{23}\cos\theta_{13})}{r_{13}} \frac{\partial{\mathcal C}_{13}}{\partial (\cos\theta_{13})} \cr
&&  \hspace{320pt} + \frac{1}{2} (1+\cos^2\theta_{12}) \frac{\partial{\mathcal C}_{13}}{\partial Z_{13}} \Bigg\}
\end{eqnarray} 
\begin{eqnarray}\lb{term1} 
 && F_{132} = \frac{k_BT}{4\pi\eta r_{13}} \bar{c}'(z_1) \Bigg\{ -(\cos\theta_{23}	+\cos\psi_{12} \cos\theta_{13}) \frac{\partial{\mathcal C}_{23}}{\partial r_{23}} 
 - \frac{\sin^2\theta_{23}+\cos\theta_{13}(\cos\theta_{13}-\cos\psi_{12}\cos\theta_{23})}{r_{23}} \frac{\partial{\mathcal C}_{23}}{\partial (\cos\theta_{23})} \cr
&&  \hspace{320pt} + \frac{1}{2} (1+\cos^2\theta_{13}) \frac{\partial{\mathcal C}_{23}}{\partial Z_{23}} \Bigg\}
\end{eqnarray} 
\begin{eqnarray}\lb{term1} 
 && F_{312} = \frac{k_BT}{4\pi\eta r_{12}} \bar{c}'(z_3) \Bigg\{ (\cos\theta_{12}	+\cos\psi_{23} \cos\theta_{13}) \frac{\partial{\mathcal C}_{12}}{\partial r_{12}} 
 + \frac{\sin^2\theta_{12}+\cos\theta_{13}(\cos\theta_{13}-\cos\psi_{23}\cos\theta_{23})}{r_{12}} \frac{\partial{\mathcal C}_{12}}{\partial (\cos\theta_{12})} \cr
&&  \hspace{320pt} + \frac{1}{2} (1+\cos^2\theta_{13}) \frac{\partial{\mathcal C}_{12}}{\partial Z_{12}} \Bigg\}
\end{eqnarray} 
\begin{eqnarray}\lb{term1} 
 && F_{231} = \frac{k_BT}{4\pi\eta r_{23}} \bar{c}'(z_2) \Bigg\{ -(\cos\theta_{13}	+\cos\psi_{12} \cos\theta_{23}) \frac{\partial{\mathcal C}_{13}}{\partial r_{13}} 
 - \frac{\sin^2\theta_{13}+\cos\theta_{23}(\cos\theta_{23}-\cos\psi_{12}\cos\theta_{13})}{r_{13}} \frac{\partial{\mathcal C}_{13}}{\partial (\cos\theta_{13})} \cr
&&  \hspace{320pt} + \frac{1}{2} (1+\cos^2\theta_{23}) \frac{\partial{\mathcal C}_{13}}{\partial Z_{13}} \Bigg\}
\end{eqnarray} 
\begin{eqnarray}\lb{term1} 
 && F_{321} = \frac{k_BT}{4\pi\eta r_{23}} \bar{c}'(z_3) \Bigg\{ (-\cos\theta_{12}	+\cos\psi_{13} \cos\theta_{23}) \frac{\partial{\mathcal C}_{12}}{\partial r_{12}} 
 - \frac{\sin^2\theta_{12}+\cos\theta_{23}(\cos\theta_{23}+\cos\psi_{13}\cos\theta_{12})}{r_{12}} \frac{\partial{\mathcal C}_{12}}{\partial (\cos\theta_{12})} \cr
&&  \hspace{320pt} + \frac{1}{2} (1+\cos^2\theta_{23}) \frac{\partial{\mathcal C}_{12}}{\partial Z_{12}} \Bigg\}
\end{eqnarray} 

\newpage

In order to calculate the leading-order asymptotic approximation to ${\mathcal C}_{123}$ for short times and large-$r,$
we must substitute into the forcing terms the corresponding derivatives of the mean,  
$\nabla \bar{c}(z,t)=c_0 \exp(-z^2/4D(t+\tau))/\sqrt{4\pi D(t+\tau)},$ and of the second cumulants. 
The latter are obtained from equation (6) in the Main Text to be 
\begin{eqnarray}\lb{rderC}
\frac{\partial{\mathcal C}_{pq}}{\partial r_{pq}} =  -  \frac{{\mathcal C}_{pq}}{r_{pq}} 
 -  \frac{3 c_0^2}{16\pi} \frac{\sigma}{Z_{pq}^2+ \frac{1}{4}z_{pq}^2}(1+\cos^2\theta_{pq}) \cos^2\theta_{pq} 
\left\{ \exp\left( -\frac{Z_{pq}^2+ \frac{1}{4}z_{pq}^2}{2D(t+\tau) }\right)  - \exp\left( -\frac{Z_{pq}^2+ \frac{1}{4}z_{pq}^2}{2D\tau }\right) \right\} \lb{dC2dr} 
\end{eqnarray} 
\begin{eqnarray}\lb{costhderC}
&&\frac{\partial{\mathcal C}_{pq}}{\partial (\cos\theta_{pq})} =  -  \frac{2\cos\theta_{pq}}{ 1+ \cos^2\theta_{pq}}{\mathcal C}_{pq} 
 -  \frac{3 c_0^2}{16\pi} \frac{\sigma r_{pq}}{Z_{pq}^2+ \frac{1}{4}z_{pq}^2}(1+\cos^2\theta_{pq}) \cos\theta_{pq} 
 \left\{ \exp\left( -\frac{Z_{pq}^2+ \frac{1}{4}z_{pq}^2}{2D(t+\tau) }\right)  - \exp\left( -\frac{Z_{pq}^2+ \frac{1}{4}z_{pq}^2}{2D\tau }\right) \right\} \cr
&& \lb{cC2dcosth} 
\end{eqnarray} 
\begin{eqnarray}\lb{ZderC} 
\frac{\partial{\mathcal C}_{pq}}{\partial Z_{pq}} =   
-  \frac{3 c_0^2}{4\pi} \frac{\sigma}{r_{pq}} (1+\cos^2\theta_{pq}) \frac{Z_{pq}}{Z_{pq}^2+ \frac{1}{4}z_{pq}^2} 
\left\{ \exp\left( -\frac{Z_{pq}^2+ \frac{1}{4}z_{pq}^2}{2D(t+\tau) }\right)  - \exp\left( -\frac{Z_{pq}^2+ \frac{1}{4}z_{pq}^2}{2D\tau }\right) \right\} \lb{dC2dZ}
\end{eqnarray} 

Finally, we must integrate in time the six forcing terms and sum them to obtain the asymptotic 
result for ${\mathcal C}_{123}.$ The time-integration of the products of the mean gradient and of the derivatives 
of the second cumulants in \eqref{dC2dr}-\eqref{dC2dZ} leads, in addition to elementary integrals, also to two less 
trivial integrals. The first can be evaluated by the change of variables $u=1/(s+\tau)$ as follows 
\be  \int_0^t \frac{e^{-c/(s+\tau)}}{\sqrt{s+\tau}} \, ds 
=c^{1/2} \left[\Gamma\left(-\frac{1}{2},\frac{c}{t+\tau}\right)-\Gamma\left(-\frac{1}{2},\frac{c}{\tau}\right)\right]\ee
where $\Gamma(\alpha,z)$ is the incomplete Gamma function 
\cite[\href{https://dlmf.nist.gov/8.2.E2}{(8.2.2)}]{NIST:DLMF}. Thus, one gets the first integral 
\begin{eqnarray}\lb{int1}  
&& I_1= \int_0^t \frac{e^{-a/(s+\tau)}}{\sqrt{s+\tau}} \left[e^{-b/(s+\tau)}-e^{-b/\tau}  \right]\, ds \cr
&& = (a+b)^{1/2} \Bigg[  \Gamma\left(-\frac{1}{2},\frac{a+b}{t+\tau} \right)  
- \Gamma\left(-\frac{1}{2},\frac{a+b}{\tau}\right) \Bigg] 
 - a^{1/2} e^{-\frac{b}{\tau}}\Bigg[  \Gamma\left(-\frac{1}{2},\frac{a}{t+\tau}\right)  
- \Gamma\left(-\frac{1}{2},\frac{a}{\tau}\right) \Bigg].  
\end{eqnarray} 
By using the expression for the incomplete Gamma function in terms of the 
Kummer confluent hypergeometric function $U$ 
\cite[\href{https://dlmf.nist.gov/8.5.E3}{(8.5.3)}]{NIST:DLMF}: 
\be \Gamma\left(-\frac{1}{2},z\right)= e^{-z} U\left(\frac{3}{2}, \frac{3}{2}; z\right), \ee  
one thus gets also 
\begin{eqnarray}\lb{int1U}  
&& I_1= \int_0^t \frac{e^{-a/(s+\tau)}}{\sqrt{s+\tau}} \left[e^{-b/(s+\tau)}-e^{-b/\tau}  \right]\, ds \cr
&& = (a+b)^{1/2} \Bigg[  e^{-\frac{a+b}{t+\tau}}U\left(\frac{3}{2}, \frac{3}{2};\frac{a+b}{t+\tau}\right) 
- e^{-\frac{a+b}{\tau}}U\left(\frac{3}{2}, \frac{3}{2};\frac{a+b}{\tau}\right) \Bigg]
- a^{1/2} e^{-\frac{b}{\tau}}\Bigg[  e^{-\frac{a}{t+\tau}}U\left(\frac{3}{2}, \frac{3}{2};\frac{a}{t+\tau}\right) 
    - e^{-\frac{a}{\tau}}U\left(\frac{3}{2}, \frac{3}{2};\frac{a}{\tau}\right) \Bigg]. \cr
&&    
\end{eqnarray}  
The second time integral that appears is 
\be 
I_2= \int_0^t \frac{e^{-a/(s+\tau)}}{\sqrt{s+\tau}} \left[E_1\left(\frac{b}{s+\tau}\right)- E_1\left(\frac{b}{\tau}\right)\right]\, ds 
\lb{int2} \ee 
and we have found no exact analytical expression for it. Instead we shall evaluate $I_2$ by numerical quadrature. 

For short times $t,$ the analytical results \eqref{int1},\eqref{int1U} for $I_1$ and the integrand in 
the definition \eqref{int2} of $I_2$ are all subject to cancellations from subtracting nearly equal terms.
In that case we can obtain analytical results from series expansion. For $I_1$ we can use the series representation 
of the incomplete Gamma function 
\cite[\href{https://dlmf.nist.gov/8.7.E3}{(8.7.3)}]{NIST:DLMF} and the binomial expansion to obtain 
\bea \lb{Gamexp} 
c^\alpha \left[\Gamma\left(-\alpha,\frac{c}{t+\tau}\right) - \Gamma\left(-\alpha,\frac{c}{\tau}\right) \right] 
= \tau^{\alpha} \sum_{n=0}^\infty \frac{(-1)^n}{n!} 
\left(\frac{c}{\tau}\right)^n\sum_{k=1}^\infty (-1)^{k-1}
\frac{(n-\alpha+1)^{k+1}}{k!} \left(\frac{t}{\tau}\right)^k \cr
&&
\eea 
We apply this for $\alpha=1/2$ with $c=a+b,\, a.$ For the integrand of $I_2$ we 
use \cite[\href{https://dlmf.nist.gov/6.6.E2}{(6.6.2)}]{NIST:DLMF} to obtain 
\bea\lb{int2exp}
E_1\left(\frac{b}{s+\tau}\right)- E_1\left(\frac{b}{\tau}\right) 
 = \ln\left(1+\frac{s}{\tau}\right) -\sum_{k=1}^\infty \frac{1}{k} \left( \frac{-b}{s+\tau}\right)^k 
\sum_{j=1}^k \frac{1}{j!(k-j)!} \left(\frac{s}{\tau}\right)^j \cr
&&
\eea 
When $\frac{a+b}{\tau}$ is sufficiently large, above some threshold $\Theta,$ then the expression 
\eqref{int1U} for $I_1$ is accurate. In our numerical work, we took $\Theta=10.$ However, 
for $\frac{a+b}{\tau}<\Theta$ we must use  the expansion \eqref{Gamexp} for $\alpha=1/2$ with $c=a+b,\, a.$ 
There is still a difficulty, because the factor $(a+b)^n-a^n e^{-b/\tau}$ in the double summation is subject 
to loss of significance error when $b$ is small. Thus, we rewrite this term as
\bea 
&& (a+b)^n-a^n e^{-b/\tau} = 2 a^n e^{-b/2\tau} \sinh\left(\frac{b}{2\tau}\right) 
 +\sum_{m=1}^n \binom{n}{m} a^{n-m} b^m \eea  
to avoid the cancellation of significant digits. For $\frac{b}{\tau}$ sufficiently large, 
above some threshold $\Theta,$ then the expression for $I_2$ in \eqref{int2} for the integrand is accurate. 
We took $\Theta=1.$ Otherwise, we define the integral by the expansion \eqref{int2exp}. In either case, 
the time-integral $I_2$ is evaluated using the 2nd-order composite trapezoid rule. A script in {\tt Matlab}  
is appended below, which carries out all of the numerical evaluations discussed above. 

Finally, we note that great simplifications occur in the large-$\tau$ limit. To leader-order 
accuracy, only terms $\propto t/\tau$ need be retained in the summations \eqref{Gamexp}-\eqref{int2exp}. 
Note, in particular, that the mean concentration gradient becomes $t$-independent to leading order:
$\nabla \bar{c} \sim \nabla c_0\sim c_0/\sqrt{4\pi D\tau}.$ The time-integrals of the forcing terms thus
simplify and reduce to time-integrals of only the derivatives of the second cumulants. These are 
easily evaluated for large-$\tau$ to be 
\bea 
 \int_0^t \frac{\partial {\mathcal C}}{\partial r_{pq}}(r_{pq},\cos\theta_{pq},Z_{pq},s)\, ds 
\sim -\frac{3}{4}|\nabla c_0|^2 \frac{\sigma}{r^2_{pq}}(1+\cos^2\theta_{pq}) D t^2\eea
\bea 
&& \int_0^t \frac{\partial {\mathcal C}}{\partial (\cos\theta_{pq})}(r_{pq},\cos\theta_{pq},Z_{pq},s)\, ds 
 \sim \frac{3}{2}|\nabla c_0|^2 \frac{\sigma}{r_{pq}} \cos\theta_{pq} D t^2\eea
\bea 
&& \int_0^t \frac{\partial {\mathcal C}}{\partial Z_{pq}}(r_{pq},\cos\theta_{pq},Z_{pq},s)\, ds  \sim -3\pi |\nabla c_0|^4 \frac{\sigma}{r_{pq}}(1+\cos^2\theta_{pq}) Z_{pq} \frac{D t^2}{c_0^2}\eea
The latter derivative is thus sub-leading compared with the other two and may be neglected. 

\begin{figure}
\centering
\includegraphics[width=0.4\textwidth]{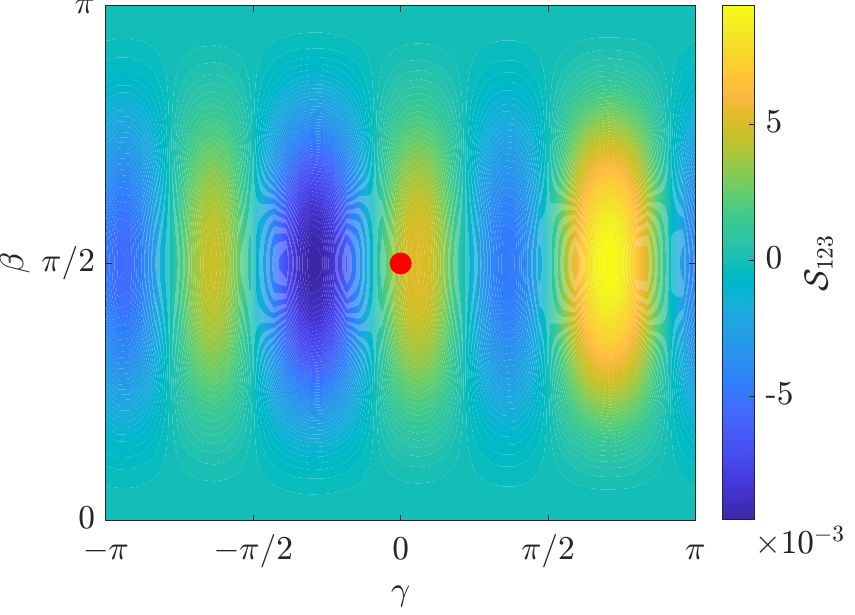}
\hfill
\includegraphics[width=0.4\textwidth]{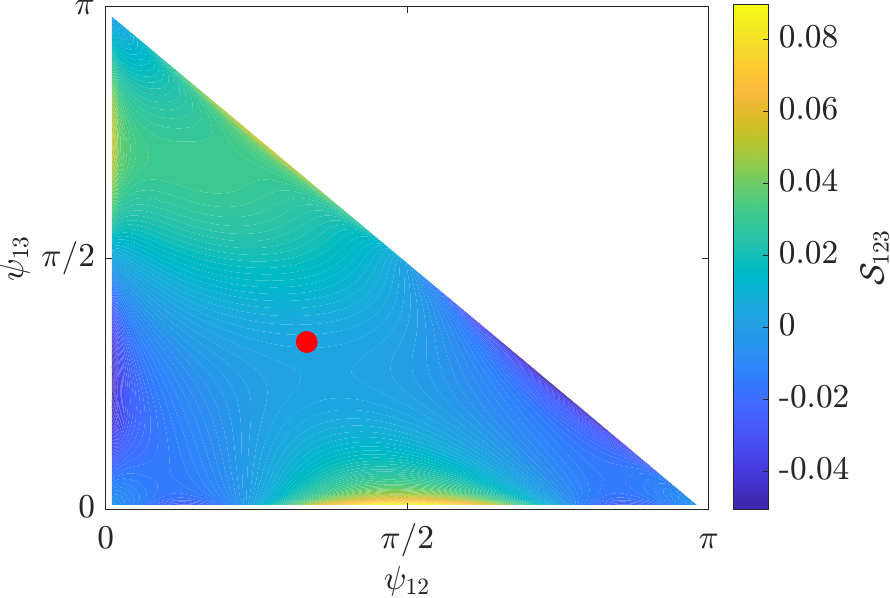}
\caption{\footnotesize {Dependence of the asymptotic three-point skewness, $\mathcal{S}_{123}=\mathcal{C}_{123}/(\mathcal{C}_{12} \mathcal{C}_{13} \mathcal{C}_{23})^{1/2}$, on 
triangle orientation (specified by Euler angles, $\beta$ and $\gamma$) and on triangle shape (specified by internal angles, $\psi_{12}$ and $\psi_{13},$ subject to the restriction $0\leq \psi_{13}+\psi_{12}\leq \pi$). In the \textit{left} panel the triangle is equilateral with side lengths $r^*=50$, while the \textit{right} panel refers to a vertical triangle, $\beta=\pi/2$ and $\gamma=0$, with arithmetic mean of side lengths $=50$. All data refer to $t^*=100$ and red dots represent the configuration studied in the Main Text.}} 
\label{skewness}
\end{figure}

To give some intuition about the complex formulas, we plot in Figure~\ref{skewness}
the three-point skewness, $\mathcal{S}_{123}=\mathcal{C}_{123}/(\mathcal{C}_{12} \mathcal{C}_{13} \mathcal{C}_{23})^{1/2}$ in the limit of vanishing mean concentration gradient, for varying orientation 
of the triangle (left panel) and for varying triangle shape (right panel). It is notable that the configuration studied in the main text does not have maximum possible skewness.

\clearpage 

\begin{lstlisting}
clear all

% INPUTS

% geometry

r13=1e4;
r23=1e4;
ps12=3*pi/2;

% orientation 

% alp=0, gam from 0 to 2*pi, bet from 0 to pi
alp=0;
bet=pi/2;
gam=pi/3; 

% vertical location
z3=-5e3; 

% time interval and time step 

T=1;
dt=0.005;
t=0:dt:T; 

% gradient parameter

tau=1e13; 


% OUTPUTS 

% derived quantities for triangle 

r12=sqrt(r13^2+r23^2-2*r13*r23*cos(ps12)); 
ps23=asin(r23*sin(ps12)/r12);
ps13=asin(r13*sin(ps12)/r12);

costh13=sin(bet)*sin(gam);
costh23=sin(bet)*sin(gam+ps12);
costh12=sin(bet)*sin(gam-ps23); 


lh=[cos(alp)*cos(gam)-sin(alp)*cos(bet)*sin(gam); ...
    sin(alp)*cos(gam)+cos(alp)*cos(bet)*sin(gam); ...
    sin(bet)*sin(gam)]; 

mh=[-cos(alp)*sin(gam)-sin(alp)*cos(bet)*cos(gam); ...
    -sin(alp)*sin(gam)+cos(alp)*cos(bet)*cos(gam); ...
    sin(bet)*cos(gam)]; 

nh=[sin(bet)*sin(alp); -sin(bet)*cos(alp); cos(bet)];

rv13=r13*lh;
rv23=r23*(cos(ps12)*lh+sin(ps12)*mh);
rv12=r12*(cos(ps23)*lh-sin(ps23)*mh);

z1=z3+rv13(3); z2=z3+rv23(3);
Z12=(z1+z2)/2; Z23=(z2+z3)/2; Z13=(z1+z3)/2;

% FIRST MOMENTS 

C1=meanf(z1,tau,t); 
C2=meanf(z2,tau,t);
C3=meanf(z3,tau,t);

% SECOND MOMENTS 

C13=covf(r13,costh13,Z13,tau,t); 
C23=covf(r23,costh23,Z23,tau,t); 
C12=covf(r12,costh12,Z12,tau,t); 

% THIRD MOMENT 

btaumax=0; 


% source term 123
A=costh23-cos(ps13)*costh12;
B=(1-costh23^2+costh12*(costh12+cos(ps13)*costh23))/r23;
C=(1+costh12^2)/2; 

a=z1^2/2;
z23=r23*costh23; 
b=Z23^2+z23^2/4; btaumax=max(b/tau,btaumax); 
I1=int1(a,b,tau,t); 
I2=int2(a,b,tau,dt,T); 

dC123=-A*(3*(1+costh23^2)/r23^2/8/pi)*I2; 
dC123= dC123 -A*(3*(1+costh23^2)*costh23^2/b/16/pi)*I1; 
dC123= dC123 + B*(3*costh23/r23/4/pi)*I2; 
dC123= dC123 - B*(3*r23*(1+costh23^2)*costh23/b/16/pi)*I1;
dC123= dC123 -C*(3*(1+costh23^2)*Z23/b/r23/4/pi)*I1; 
dC123=dC123/r12;

C123=dC123; 

% source term 213
A=costh13+cos(ps23)*costh12;
B=(1-costh13^2+costh12*(costh12-cos(ps23)*costh13))/r13;
C=(1+costh12^2)/2; 

a=z2^2/2;
z13=r13*costh13; 
b=Z13^2+z13^2/4; btaumax=max(b/tau,btaumax); 
I1=int1(a,b,tau,t);
I2=int2(a,b,tau,dt,T);

dC123=-A*(3*(1+costh13^2)/r13^2/8/pi)*I2; 
dC123= dC123 -A*(3*(1+costh13^2)*costh13^2/b/16/pi)*I1; 
dC123= dC123 + B*(3*costh13/r13/4/pi)*I2; 
dC123= dC123 - B*(3*r13*(1+costh13^2)*costh13/b/16/pi)*I1;
dC123= dC123 -C*(3*(1+costh13^2)*Z13/b/r13/4/pi)*I1; 
dC123=dC123/r12;

C123=C123+dC123; 

% source term 132
A=-(costh23+cos(ps12)*costh13);
B=-(1-costh23^2+costh13*(costh13-cos(ps12)*costh23))/r23;
C=(1+costh13^2)/2; 

a=z1^2/2;
z23=r23*costh23; 
b=Z23^2+z23^2/4; btaumax=max(b/tau,btaumax); 
I1=int1(a,b,tau,t);
I2=int2(a,b,tau,dt,T);

dC123=-A*(3*(1+costh23^2)/r23^2/8/pi)*I2; 
dC123= dC123 -A*(3*(1+costh23^2)*costh23^2/b/16/pi)*I1; 
dC123= dC123 + B*(3*costh23/r23/4/pi)*I2; 
dC123= dC123 - B*(3*r23*(1+costh23^2)*costh23/b/16/pi)*I1;
dC123= dC123 -C*(3*(1+costh23^2)*Z23/b/r23/4/pi)*I1; 
dC123=dC123/r13;

C123=C123+dC123; 

% source term 312
A=costh12+cos(ps23)*costh13;
B=(1-costh12^2+costh13*(costh13-cos(ps23)*costh23))/r12;
C=(1+costh13^2)/2; 

a=z3^2/2;
z12=r12*costh12; 
b=Z12^2+z12^2/4; btaumax=max(b/tau,btaumax); 
I1=int1(a,b,tau,t);
I2=int2(a,b,tau,dt,T);

dC123=-A*(3*(1+costh12^2)/r12^2/8/pi)*I2; 
dC123= dC123 -A*(3*(1+costh12^2)*costh12^2/b/16/pi)*I1; 
dC123= dC123 + B*(3*costh12/r12/4/pi)*I2; 
dC123= dC123 - B*(3*r12*(1+costh12^2)*costh12/b/16/pi)*I1;
dC123= dC123 -C*(3*(1+costh12^2)*Z12/b/r12/4/pi)*I1; 
dC123=dC123/r13;

C123=C123+dC123; 

% source term 231
A=-(costh13+cos(ps12)*costh23);
B=-(1-costh13^2+costh23*(costh23-cos(ps12)*costh13))/r13;
C=(1+costh23^2)/2; 

a=z2^2/2;
z13=r13*costh13; 
b=Z13^2+z13^2/4; btaumax=max(b/tau,btaumax); 
I1=int1(a,b,tau,t);
I2=int2(a,b,tau,dt,T);

dC123=-A*(3*(1+costh13^2)/r13^2/8/pi)*I2; 
dC123= dC123 -A*(3*(1+costh13^2)*costh13^2/b/16/pi)*I1; 
dC123= dC123 + B*(3*costh13/r13/4/pi)*I2; 
dC123= dC123 - B*(3*r13*(1+costh13^2)*costh13/b/16/pi)*I1;
dC123= dC123 -C*(3*(1+costh13^2)*Z13/b/r13/4/pi)*I1; 
dC123=dC123/r23;

C123=C123+dC123;

% source term 321
A=-costh12+cos(ps13)*costh23;
B=-(1-costh12^2+costh23*(costh23+cos(ps13)*costh12))/r12;
C=(1+costh23^2)/2; 

a=z3^2/2;
z12=r12*costh12; 
b=Z12^2+z12^2/4; btaumax=max(b/tau,btaumax); 
I1=int1(a,b,tau,t);
I2=int2(a,b,tau,dt,T);

dC123=-A*(3*(1+costh12^2)/r12^2/8/pi)*I2; 
dC123= dC123 -A*(3*(1+costh12^2)*costh12^2/b/16/pi)*I1; 
dC123= dC123 + B*(3*costh12/r12/4/pi)*I2; 
dC123= dC123 - B*(3*r12*(1+costh12^2)*costh12/b/16/pi)*I1;
dC123= dC123 -C*(3*(1+costh12^2)*Z12/b/r12/4/pi)*I1; 
dC123=dC123/r23;

C123=C123+dC123; 


C123=C123*.75/sqrt(2*pi); 

figure
plot(t,C123)
title('triple cumulant')

figure 
N123=C123./sqrt(abs(C13.*C23.*C12));
plot(t,N123)
title('normalized triple cumulant')

btaumax=btaumax


function y=int1(a,b,tau,t)

if a+b<=10*tau

y=zeros(size(t));
nn=0; 
next=ones(size(t)); 
fac=1; 
while norm(next,'inf')>eps*norm(y,'inf')
next=t/tau; 
dnext=next;
kk=2;
while norm(dnext,'inf')>eps*norm(next,'inf')
dnext=-(nn+kk-1.5)*dnext.*(t/tau)/kk;
next=next+dnext;
kk=kk+1;
end      
mult=a^nn*2*exp(-b/tau/2)*sinh(b/tau/2);
if nn>=1
for mm=1:nn    
mult=mult+nchoosek(nn,mm)*a^(nn-mm)*b^mm; 
end 
end     
next=fac*next*mult; 
y=y+next; 
nn=nn+1;
fac=-fac/nn/tau;
end 
y=sqrt(tau)*y; 

else     
y=sqrt(a+b)*(exp(-(a+b)./(t+tau)).*kummerU(1.5,1.5,(a+b)./(t+tau))...
             -exp(-(a+b)/tau)*kummerU(1.5,1.5,(a+b)/tau));
y=y-sqrt(a)*exp(-b/tau)*(exp(-a./(t+tau)).*kummerU(1.5,1.5,a./(t+tau))...
             -exp(-a/tau)*kummerU(1.5,1.5,a/tau)); 
end 

end

function y=int2(a,b,tau,dt,tf)

s=0:dt:tf;

if b<=5*tau 
diff=log(1+s/tau);  
next=1;
kk=1; 
while norm(next,'inf')>eps*norm(diff,'inf')
jj=1:kk;
S=ones(kk,1)*s;
JJ=jj.'*ones(size(s)); 
nsum=sum((S/tau).^JJ./factorial(JJ)./factorial(kk-JJ),1);
next=nsum.*(-b./(s+tau)).^kk/kk; 
diff=diff-next;
kk=kk+1; 
end     
f=exp(-a./(s+tau)).*diff./sqrt(s+tau);

else 
f=exp(-a./(s+tau)).*(expint(b./(s+tau))-expint(b/tau))./sqrt(s+tau);
end 
y=cumtrapz(s,f); 

end

function y=meanf(z,tau,t)

y=(1+erf(z./sqrt(2*(t+tau))))/2;

end 

function y=covf(r,costh,Z,tau,t)

z=r*costh;
b=Z^2+z^2/4;

if b<= tau 
diff=log(1+t/tau);  
next=ones(size(t));
kk=1; 
while norm(next,'inf')>eps*norm(diff,'inf')
jj=1:kk;
T=ones(kk,1)*t;
JJ=jj.'*ones(size(t)); 
nsum=sum((T/tau).^JJ./factorial(JJ)./factorial(kk-JJ),1);
next=nsum.*(-b./(t+tau)).^kk/kk; 
diff=diff-next;
kk=kk+1; 
end     
y=3*(1+costh^2)*diff/r/(8*pi);

else     
y=3*(1+costh^2)*(expint(b./(t+tau))-expint(b/tau))/r/(8*pi);
end 

end


\end{lstlisting}

\newpage 

\section{Lagrangian Monte Carlo Numerical Method and HPC implementation}

We exploit a Lagrangian numerical scheme that was developed to calculate statistical 
correlation functions of a passive scalar advected by the Kraichnan model of a turbulent velocity field
\cite{frisch1998intermittency,frisch1999lagrangian,gat1998anomalous}. 
To calculate the $P$th moment function of the concentration field 
$\langle c({\bf x}_1,t)\cdot\cdot\cdot\cdot \ c({\bf x}_P,t)\rangle,$ 
the DFV model is first solved for $c({\bf x}_p,t)$ at the points $\bx_p,$ $p=1,...,P$
in one realization of the thermal velocity $\bw$. This is accomplished by setting 
$c({\bf x}_p,t)=c_0({\boldsymbol \xi}_p(t))$
where $d{\boldsymbol \xi}_p/dt={\bf w}_p,$ ${\boldsymbol \xi}_p(0)=\bx_p,$ $p=1,...,P$ with 
${\boldsymbol {\mathcal W}}=({\bf w}_1,...,{\bf w}_P)$ a $3P$-dimensional Gaussian white-noise with covariance 
\begin{equation}
   \langle w_{pi}(t)w_{qj}(t')\rangle_{\boldsymbol\Xi}=R_{pq}({\boldsymbol \xi}_p(t),{\boldsymbol \xi}_q(t))\delta(t-t')
   \,,
   \label{eq:R}
\end{equation}
for fixed ${\boldsymbol \Xi}=({\boldsymbol \xi}_1,...,{\boldsymbol \xi}_P)$ and 
for $p,q = 1,...,P$ and $i,j=1,...,d$ in space dimension $d.$ The evolution of the $P$ stochastic Lagrangian tracers, 
${\boldsymbol \xi}_p(t)$, $p=1,...,P$ is implemented with an Euler-Maruyama discretization
\begin{equation}
   {\boldsymbol \xi}_p(t+\Delta t)={\boldsymbol \xi}_p(t)+\tilde{\bf w}_p\sqrt{\Delta t}, \quad p=1,...,P,
    \label{eq:EM}
\end{equation}
where $\tilde{\boldsymbol {\mathcal W}}_P=(\tilde {\bf w}_1,...,\tilde {\bf w}_P)$ is a normal random 
variable having mean zero and $Pd\times Pd$ covariance matrix
\begin{equation}
   R_{pi,qj}=R_{pq}({\boldsymbol \xi}_p(t),{\boldsymbol \xi}_q(t))
   \,,
   \label{eq:R}
\end{equation}
for $p,q = 1,...,P$ and $i,j=1,...,d$ in $d$ space dimensions. We can generate an 
independent sample of the $Pd$-random vector by writing it as 
$\tilde{\boldsymbol {\mathcal W}}_P={\bf L}_P \tilde {\bf N}_P$, where ${\bf L}_P$ 
is the lower triangular Cholesky factor of the positive-definite 
covariance matrix \eqref{eq:R} and $\tilde {\bf N}_P$ is a $Pd$-dimensional 
standard normal vector. Since our focus is on the triple cumulant, we thus take
$P=3$ and evolve a triplet ${\boldsymbol \Xi}=({\boldsymbol \xi}_1,{\boldsymbol \xi}_2,{\boldsymbol \xi}_3)$
of stochastic Lagrangian tracers in each realization ${\bf w}$ of the thermal velocity field,
obtaining thereby $c({\bf x}_p,t)=c_0({\boldsymbol \xi}_p(t))$ for $p=1,2,3.$

As in the earlier work on turbulent advection \cite{frisch1998intermittency,frisch1999lagrangian,gat1998anomalous}, 
the ensemble average that defines the triple cumulant is then approximated by an empirical average 
\be C({\bf x}_1,{\bf x}_2,{\bf x}_3, t)\simeq \frac{1}{N} \sum_{n=1}^N 
c^{\prime(n)}({\bf x}_1,t)c^{\prime(n)}({\bf x}_2,t)c^{\prime(n)}({\bf x}_3,t) 
\lb{eq:sample} \ee 
over a large number $N$ of independent samples, where the concentration fluctuation field 
is defined by $c^{\prime(n)}({\bf x}_p,t)=c_0({\boldsymbol \xi}_p^{(n)}(t))-\bar{c}({\bf x}_p,t)$ 
and ${\boldsymbol \xi}_p^{(n)}(t),$ $n=,1,...,N$ are independent solutions 
of \eqref{eq:EM}. The 
convergence of the sample average \eqref{eq:sample} as $N\to\infty$ is slow, with 
typical Monte Carlo errors $\propto 1/\sqrt{N}.$ Nevertheless, the scheme yielded 
accurate results for turbulent advection 
\cite{frisch1998intermittency,frisch1999lagrangian,gat1998anomalous}
with a number of samples at most $N\sim 10^7.$ Unfortunately, 
our problem differs in two crucial respects. First, the covariance for the Kraichnan 
model of a turbulent velocity field had covariance increasing with $r,$ whereas the velocity 
corresponding to thermal noise has covariance proportional to the Oseen tensor 
\be R_{ij}({\bf r})=\frac{k_BT}{4\pi \eta r}(\delta_{ij}+\hat{r}_i\hat{r}_j), \quad r\gtrsim \sigma 
\lb{oseen} \ee 
and thus decreasing with $r.$ The consequence is that the elements in the 
$Pd\times Pd$ matrix \eqref{eq:R} off-diagonal in $p,q,$ which are entirely responsible 
for producing correlations between particles, grow smaller in time for our problem 
as particles separate. The second key difference is that the turbulence studies 
\cite{frisch1998intermittency,frisch1999lagrangian,gat1998anomalous} considered
a statistical steady-state for the passive scalar with large-scale injection, 
whereas we study the problem of free diffusion for which the concentration correlations decay in time.
This decaying magnitude makes it more difficult to achieve acceptable relative error. 
Both of these differences require that $N$ must be very much larger in our problem.
Fortunately, the calculation of the sample average \eqref{eq:sample} is trivially 
parallelizable, with only the requirement to choose independent seeds for pseudorandom 
number generators on different processors. 

We have developed and implemented a Monte Carlo algorithm based on a regularized version of \eqref{oseen} that accounts for the filtering of wavenumbers $\gtrsim 1/ \sigma$. Specifically, we implement the exponentially decaying kernel proposed in \cite{eyink2022high}, Appendix D, that also reproduces the expected Stokes-Einstein macroscopic diffusivity $D=k_BT/6\pi\eta \sigma.$ For the sake of better performance, the numerical solver implements the regularization from \cite{eyink2022high} in \eqref{eq:R} when $r_{pq}:=|{\boldsymbol \xi}_p-{\boldsymbol \xi}_q|\le10\sigma$ and employs \eqref{oseen} otherwise. Moreover, we exploit an adaptive time-stepping method for the Euler-Maruyama time-discretization in \eqref{eq:EM} such that $\Delta t= f \rho^2/2D$, where $\rho=\min_{p,q}|{\boldsymbol \xi}_p-{\boldsymbol \xi}_q|$ and $0<f\le1$ is a parameter controlling what fraction of the distance between the nearest two Lagrangian tracers can be covered on average with one time-step. We use $f=0.2$ in our results -- tests with smaller and larger fractions confirm the time-scheme is converging. In addition, the results of the Main Text related to the long-time behavior are obtained imposing a minimum of $10$ time-steps between consecutive data points to ensure smooth plots.

\begin{figure}
\includegraphics[trim={5 20 5 20},clip,width=0.8\columnwidth]{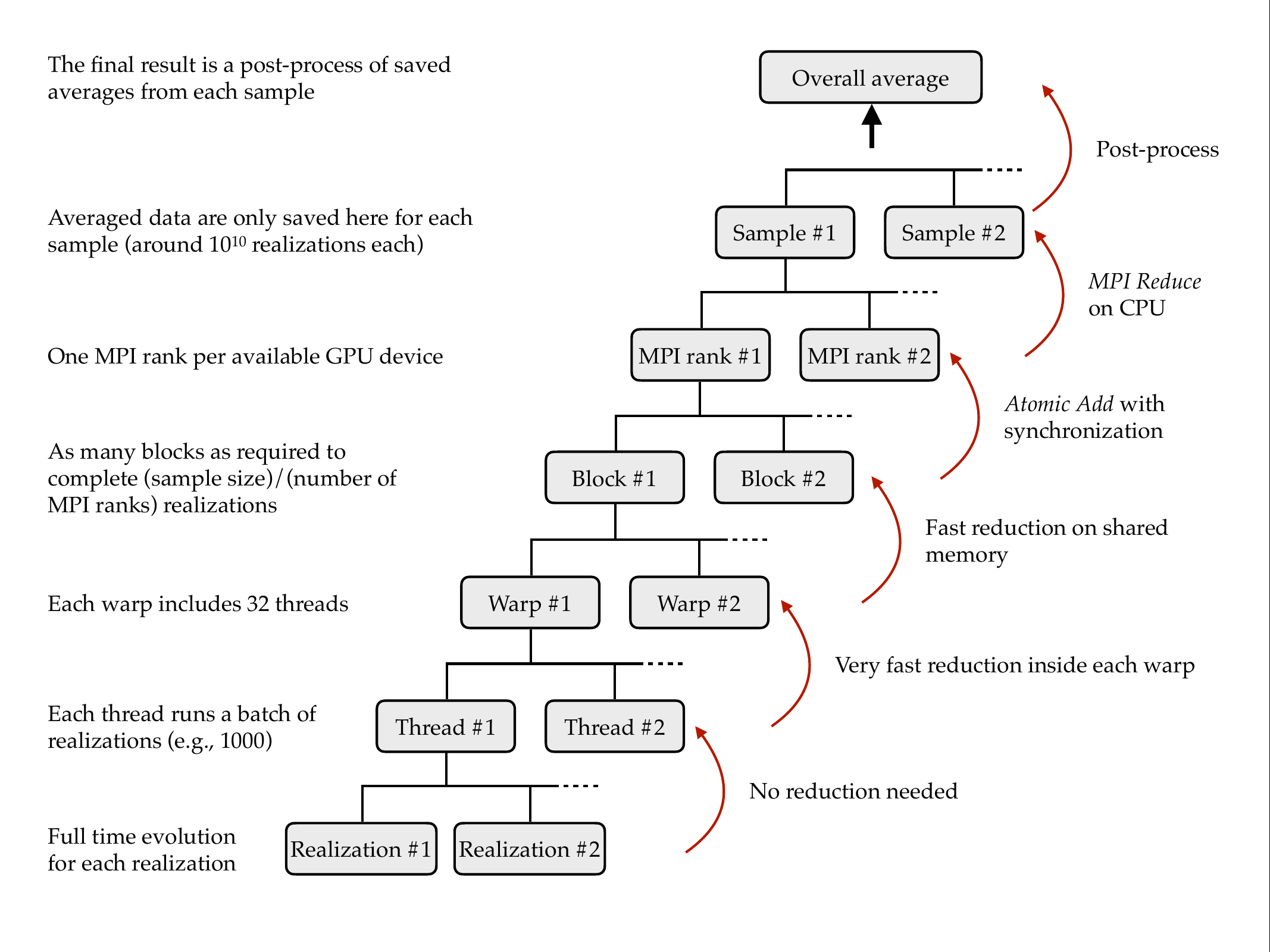}
\caption{\footnotesize {Scheme illustrating the parallelization strategy and data management. Very large samples of realizations are independently run by several MPI processes and GPU devices each. The final result is obtained through a post-processing.}} 
\label{Scheme}
\end{figure}

Although conceptually simple, the Lagrangian Monte Carlo algorithm requires careful implementation to efficiently support an extremely high number of realizations. Given the algorithm's inherently parallel structure, GPU architectures are the natural choice. To this end, we have developed a CUDA+MPI implementation that minimizes communication -- the real bottleneck in reduction-heavy algorithms -- by hierarchically distributing realizations and reductions across GPU devices, GPU blocks, warps, and, finally, batches in each thread. Figure~\ref{Scheme} shows the parallelization scheme of our GPU code.

In order to manage the enormous number of realizations contributing to the average and to allow running independent Monte Carlo campaigns, we split the computation into smaller samples and compute the overall statistical observables through a post-process. This requires storing some data. Saving the trajectory of each realization for just one of the curves in the Main Text would require $\sim 300\, \rm PB$. We drastically reduce the amount of stored data to $\sim 50\,\rm MB$ by saving only the statistically meaningful observables (\textit{i.e.} the cumulants up to third order) at the desired times for very large batches of realizations (called samples in Figure~\ref{Scheme}). Since each data file now contains the averages over a smaller sample, this data management also allows the central limit theorem to be exploited to compute the Monte Carlo error as the standard deviation of the sample means divided by the square root of the number of samples. This is sensibly faster than computing the standard deviation from each single realization.

Every time the code is launched, it iteratively computes new samples independently. Each sample is distributed over several nodes and GPU devices through standard MPI parallelization. The workload is further distributed over different GPU blocks, each containing several warps of $32$ threads. Each thread then runs a batch of $1000$ realizations, making the full time evolution of the particles trajectories the innermost loop. 
In order to compute the desired averages up to the level of samples, many reductions are needed. Proceeding upwards in the scheme of Figure~\ref{Scheme}, we reduce the partial sums computed by each thread to an intermediate sum over threads of the same warp (this is very efficient). Then, the first thread of each warp inside the same block reduces the sum again exploiting the GPU's shared memory, therefore no synchronization across blocks is needed up to here. The last reduction inside the GPU device is the \texttt{atomicAdd}, which requires blocks to synchronize in order to copy the partial sum to the CPU host. Finally, \texttt{MPI\_Reduce} is used to collect the partial sums computed by each device and to store the desired averages for each sample
at the penultimate level in Figure~\ref{Scheme}

The CUDA environment also allows using very efficient random number generators. We use \texttt{Philox\_4x32\_10} from the cuRAND library with a different seed for each MPI rank (or GPU) generated from the Linux file \texttt{/dev/urandom} and a different state for each thread. Moreover, the \texttt{curand\_normal2\_double}
function within the cuRAND library is exploited to halve the cost of producing two random numbers with double precision.
Beside managing the large amount of realizations, the implementation of the Monte Carlo algorithm also requires to deal with the loss of significance, typical of averages over very large samples. We mitigate this issue by using double precision and implementing an extended version of the Kahan–Babuška compensated summation provided by Neumaier \cite{kahan1965pracniques,neumaier1974rundungsfehleranalyse} in all summations and reductions up to the block level. Distributing the summations over many levels of parallelization also helps.

In our computation, the loss of significance is even worsened by the asymptotically decaying concentration profile, $\bar c^*(z^*,t^*)$. Indeed, when the particles are outside the interface region, the concentration is almost homogeneous and the small deviations from the asymptotic value might be lost during the evaluation of the concentration fluctuation, $c_0^*({\boldsymbol \xi}^*)-\bar c^*({\bf x}^*,t^*)$.
First of all, we shift the profile to be $\in [-0.5, 0.5]$ to balance the number of summations of positive and negative quantities. 
We use the Heaviside function for $t^*+\tau^*=0$ and a combination of error function and complementary error function otherwise. In particular, if $z^*\gtrsim \sqrt{2(t^*+\tau^*)}$ or $z^*\lesssim- \sqrt{2(t^*+\tau^*)}$, then $\bar c^*(z^*,t^*)=0.5-\epsilon$ or $\bar c^*(z^*,t^*)=-0.5+\epsilon$, respectively, with $\epsilon$ a small positive deviation. Therefore, we may have loss of significance due to the leading order terms being much larger than the small deviations. To avoid this, we exploit $\textrm{erf}(z^*)=1-\textrm{erfc}(z^*),$ $\textrm{erf}(-z^*)=-\textrm{erf}(z^*)$ 
in the formula for $\bar c^*(z^*,t^*),$
thus separating the leading order term from the remainder ($\epsilon=\frac{1}{2}\textrm{erfc}|z^*|$). Therefore, the concentration fluctuation is computed by separately subtracting the leading order terms and the remainders, similar to the compensated summations algorithm mentioned above.

Overall, the Monte Carlo code has proven capable to yield up to $N\sim 10^{15}$ independent realizations. Each curve in the insets of Figure~1 and Figure~2 of the Main Text required $\sim 10^{14}$ realizations, whereas those in the main plots (long-time behavior) were obtained with $\sim 3\times 10^{13}$ realizations. All these data were produced on the Leonardo Tier-0 cluster hosted by CINECA by performing a total of $\sim 76\,\rm EFLOP$ in double precision, corresponding to $\sim 7700$ GPU-hours. Despite the very large computation, the computational cost was as low as $\sim 1$ GPU-hour per $10^{12}$ realizations. Indeed, this efficient implementation allowed to reach a performance of $5.8\,\rm TFLOP/s$ per GPU (which is close to the cluster's peak performance) and run simulations at a rate of $\sim 130\,\rm TFLOP/s$.

\newpage 

\section{Supplemental numerical data}

We provide here some further numerical data to supplement that presented in the Main Text and End Matter. 

\subsection*{Skewness for $r^*=50,$ including $\tau^*=10^4$}

We elected in Fig.~2 of the Main Text to plot numerical results for the three-point skewness
only at the three largest values $\tau^*=10^6$, $10^8,$ $10^{10},$ in order to show better 
their convergence as $\tau^*\to\infty.$ For completeness, we plot here in Fig.~\ref{S r=50}
the entire set of results including also $\tau^*=10^4$. As expected from the sign of the triple 
cumulant in Fig.~1 of the Main Text for the corresponding value of $\tau^*$, the three-point skewness
for $\tau^*=10^4$ is negative over the entire time range. A simple physical explanation for 
this sign is lacking and also for its reversal with increasing $\tau^*.$ Nevertheless, just 
as for the other three choices of $\tau^*,$ the three-point skewness appears to level off and 
approach a ``quasi-equilibrium'' value for $t^*\to\infty.$ An extremely slow decrease
to zero might also be suggested by the plotted results and cannot be ruled out. Perhaps
the most significant observation from Fig.~\ref{S r=50} is that the maximum amplitude 
of the skewness is two times larger (about 0.02) for $\tau^*=10^4$ than it is for the 
higher $\tau^*$ values. This has the interesting implication for experiment that sharper 
interfaces with larger initial concentration gradients may have greater non-Gaussian signatures
of the concentration fluctuations.

\subsection*{Cumulants for $r^*=100$, short-time regime}

In the Main Text and in the End Matter we presented results only for three points forming 
an equilateral triangle with side-length $r^*=50,$ but comparable results are obtained 
for other values of $r^*$ and for other triangle shapes and orientations. As representative 
of this fact, we show here numerical results for an equilateral triangle of the same orientation 
as the earlier case $(\gamma=0,\beta=\pi/2)$ but instead with $r^*=100.$ 
These results were computed only for short times $t^*\leq 100$ 
where they can be directly compared with the asymptotic analytical results derived in SM,\S A. 

\begin{figure}[h]
\includegraphics[trim={0 0 0 0},clip,width=0.4\columnwidth]{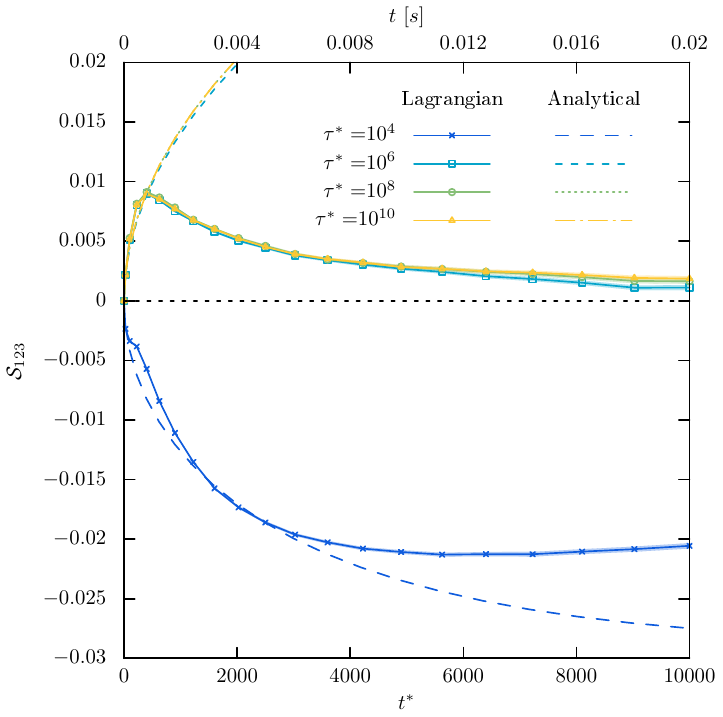}
\caption{\footnotesize {Three-point skewness for the vertical equilateral triangle with $r^*=50$ and vanishing concentration gradient, $\tau^*=10^4$, $10^6$, $10^8$, and $10^{10}$. This is an extension of Figure~2 of the Main Text. Solid lines with symbols and shades represent numerical results while dashed lines the analytical prediction.}} 
\label{S r=50}
\end{figure}

\begin{figure}[h]
\includegraphics[trim={0 0 0 0},clip,width=0.4\columnwidth]{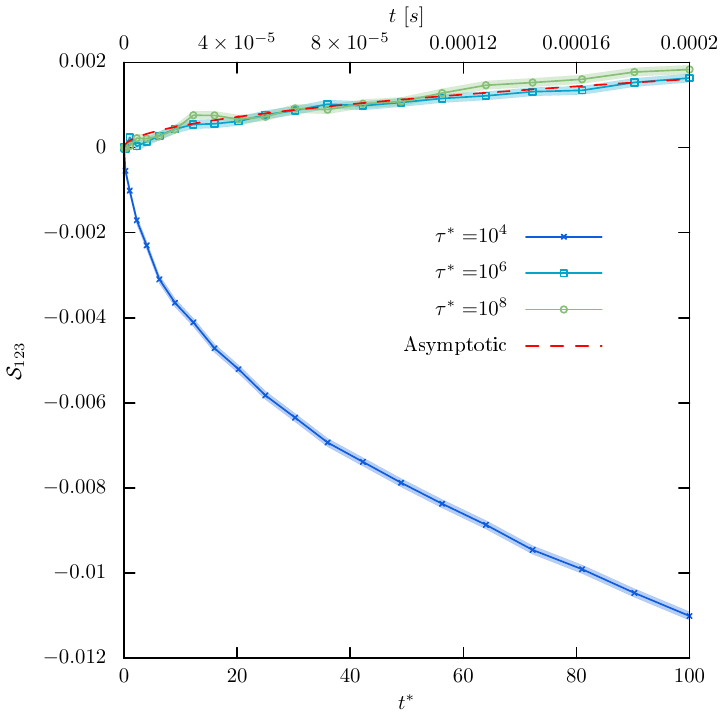}
\caption{\footnotesize {Three-point skewness for the vertical equilateral triangle with $r^*=100$ and vanishing concentration gradient, $\tau^*=10^4$, $10^6$, and $10^8$. Solid lines with symbols and shades represent numerical results while the red dashed line is the asymptotic analytical prediction in the limit of vanishing gradient.}} 
\label{S r=100}
\end{figure}

We present first in Figure \ref{S r=100} the results for the three-point skewness 
with $r^*=100,$ which can be observed to be qualitatively quite similar to those for $r^*=50$
in Fig.~\ref{S r=50}. The agreement between numerical Monte Carlo and analytical results 
is even better than for the smaller $r^*$, 
as expected since the asymptotic calculation assumes large $r^*$ and small $t^*.$
The main result of increasing $r^*$ is that the magnitudes of the three-point skewness have become smaller, 
in agreement with analytical predictions. 

\begin{figure}[h]
\includegraphics[trim={0 0 0 0},clip,width=0.5\columnwidth]{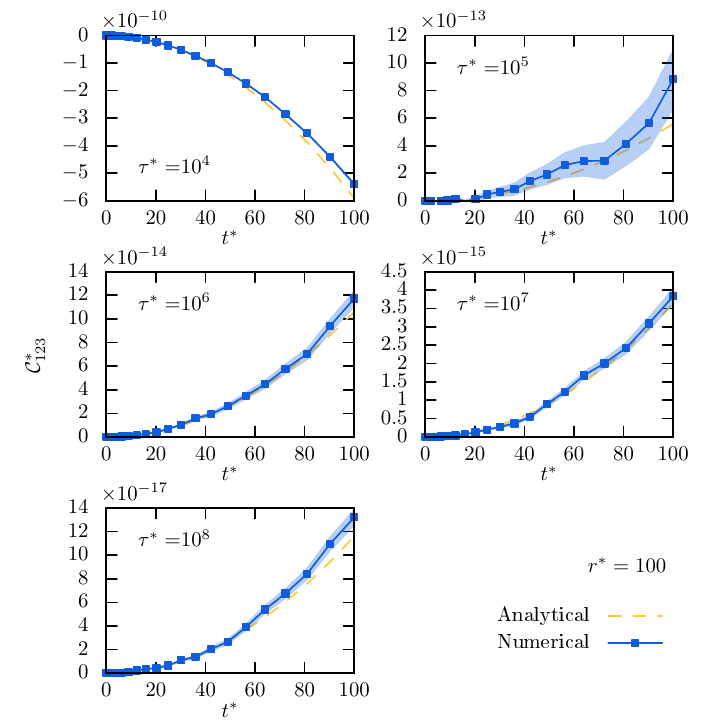}
\caption{\footnotesize {Triple cumulant for the vertical equilateral triangle with $r^*=100$ and vanishing concentration gradient, $\tau^*=10^4$, $10^5$, $10^6$, $10^7$, and $10^8$. Solid lines with symbols and shades represent numerical results while dashed lines the analytical prediction.}} 
\label{C_3 r=100}
\end{figure}

\begin{figure}[h]
\includegraphics[trim={0 0 0 0},clip,width=0.5\columnwidth]{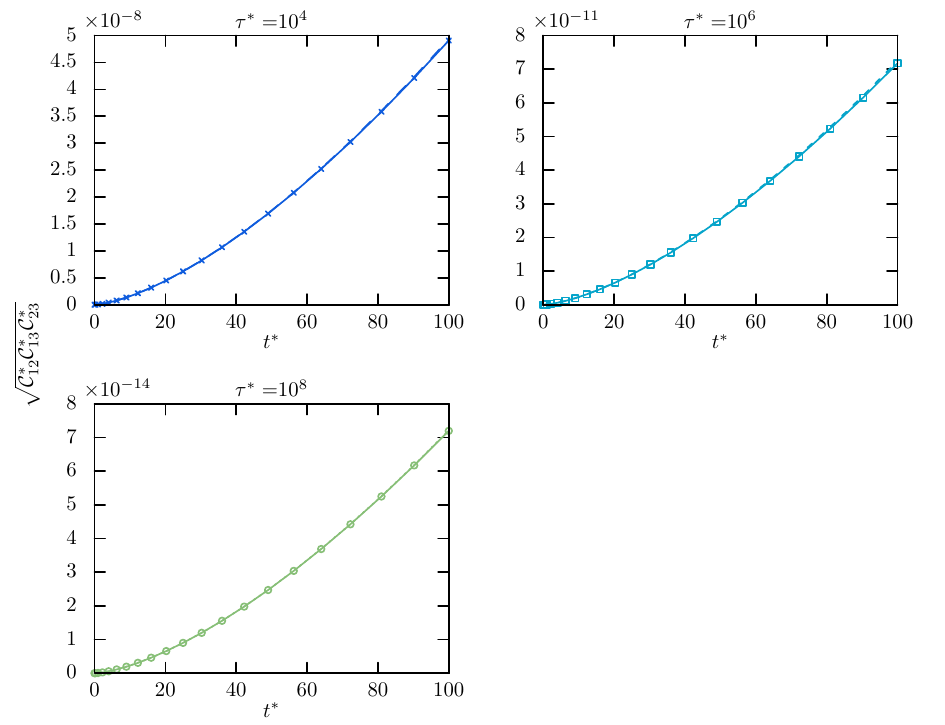}
\caption{\footnotesize {Average second cumulant for the vertical equilateral triangle with $r^*=100$ and vanishing concentration gradient, $\tau^*=10^4$, $10^6$, and $10^8$. Solid lines with symbols and shades represent numerical results while dashed lines the analytical prediction.  The Monte Carlo error is too small to be visible at this scale.}} 
\label{C_2 r=100}
\end{figure}

The input results for calculating the skewness at $r^*=100$ are presented in the remaining 
plots, Figure \ref{C_3 r=100} for the triple cumulant ${\mathcal C}_{123}$ and 
Figure \ref{C_2 r=100} for the combined second cumulants $({\mathcal C}_{13}{\mathcal C}_{23}{\mathcal C}_{12})^{1/2},$ both plotted versus $t^*.$ An interesting feature observed in Fig. \ref{C_3 r=100}
is that the Monte Carlo errors are greatest for the case $\tau^*=10^5$ at which the triple 
cumulant has switched from negative to positive. The same feature is also observed for the 
results at $r^*=50$ in Fig.~1 of the Main Text. The increase in Monte Carlo error for the same 
sample size implies an increase in the variance, which is tempting to attribute 
to the variation as realizations switch between triple products with positive and negative values.  
Note in Fig.~\ref{C_2 r=100} that the Monte Carlo error are consistently much smaller 
for the second cumulants than they are for the triple cumulants, just as was found 
for the results at $r^*=50$ in the End Matter, Fig.3.

\clearpage 

\clearpage 

\FloatBarrier
\bibliography{PRLfluctuations.bib}